\documentclass{article}

\usepackage{arxiv}
\usepackage{graphicx}
\usepackage[utf8]{inputenc} 
\usepackage[T1]{fontenc}    
\usepackage{hyperref}       
\usepackage{url}            
\usepackage{booktabs}       
\usepackage{amsfonts}       
\usepackage{nicefrac}       
\usepackage{microtype}      
\usepackage{lipsum}		
\usepackage{xcolor}
\usepackage{mathrsfs}
\usepackage{amsmath}
\usepackage{mathtools}
\usepackage{caption}
\usepackage{subcaption}
\usepackage{enumerate}
\usepackage{natbib} 
\usepackage{authblk}
\newcommand{\hs}{\mathrm{hs}}

\numberwithin{equation}{section}

\title{Latent Representation Learning of Multi-scale Thermophysics: Application to Dynamics in Shocked Porous Energetic Material}

\author[1]{Shahab Azarfar}
\author[1]{Joseph B. Choi}
\author[1]{Phong CH. Nguyen}
\author[2]{Yen T. Nguyen}
\author[2]{Pradeep Seshadri}
\author[2]{H.S. Udaykumar}
\author[1,3,*]{Stephen Baek}
\affil[1]{School of Data Science, University of Virginia, United States}
\affil[2]{Department of Mechanical Engineering, University of Iowa, United States}
\affil[3]{Department of Mechanical and Aerospace Engineering, University of Virginia, United States}
\affil[*]{Corresponding Author: baek@virginia.edu}

\begin{document}

\maketitle

\begin{abstract}
Coupling of physics across length and time scales plays an important role in the response of microstructured materials to external loads. In a multi-scale framework, unresolved (subgrid) meso-scale dynamics is upscaled to the homogenized (macro-scale) representation of the heterogeneous material through closure models. Deep learning models trained using meso-scale simulation data are now a popular route to assimilate such closure laws. However, meso-scale simulations are computationally taxing, posing practical challenges in training deep learning-based surrogate models from scratch. In this work, we investigate an alternative meta-learning approach motivated by the idea of tokenization in natural language processing. We show that one can learn a reduced representation of the micro-scale physics to accelerate the meso-scale learning process by tokenizing the meso-scale evolution of the physical fields involved in an archetypal, albeit complex, reactive dynamics problem, \textit{viz.}, shock-induced energy localization in a porous energetic material. A probabilistic latent representation of \textit{micro}-scale dynamics is learned as building blocks for \textit{meso}-scale dynamics. The \textit{meso-}scale latent dynamics model learns the correlation between neighboring building blocks by training over a small dataset of meso-scale simulations. We compare the performance of our model with a physics-aware recurrent convolutional neural network (PARC) trained only on the full meso-scale dataset. We demonstrate that our model can outperform PARC with scarce meso-scale data. The proposed approach accelerates the development of closure models by leveraging inexpensive micro-scale simulations and fast training over a small meso-scale dataset, and can be applied to a range of multi-scale modeling problems.
\end{abstract}


\section{Introduction}

Multi-scale dynamics, coupling phenomena at two or more distinct scales, arise in many physical applications, such as multi-phase flows \citep{saurel1999multiphase,udaykumar1997multiphase}, transport in porous media \citep{mcdowell1986particle,parker1989multiphase}, multi-material flows \citep{luo2004computation,banks2007high}, etc. When scales are distinctly separated, distinguished by the presence of an underlying microstructure, for example \citep{perry2018relating}, the conventional practice is to solve the dynamics at the observable (macro-) scale by homogenizing the properties and processes at the unresolved (micro-/meso-) scale. This leads to the need for closure models to connect physics across disparate scales. 

The problem of shock-induced energy localization in porous energetic material is inherently multi-scale in the following sense: at the micro-scale (\textit{i.e.}, $\mathcal{O} (1 \, \mu m)$), the collapse of isolated voids in the microstructure of an EM can create localized hotspots that may expand and lead to ignition of the surrounding material; the resulting hotspots can trigger chemical reactions that release further energy and intensify the traveling shockwave, which, eventually, can lead to a self-sustaining detonation at the macro-scale (\textit{i.e.}, $\mathcal{O} (1 \, cm)$).\footnote{In the context of porous energetic material, we refer to the length scale of a single pore (\textit{i.e.}, $\mathcal{O} (1 \, \mu m)$) and the length scale of a field of voids in a microstructure domain of size $\mathcal{O} (1 \,  mm)$ and $\mathcal{O} (1 \, cm)$ as the micro, meso and macro-scale, respectively.} For this to occur, however, the heat release from a large ensemble of hotspots must be coordinated with the overpassing shock, successively pumping energy into the shock all the way to the formation of a detonation wave, a phenomenon called shock-to-detonation transition (SDT) \citep{handley2018understanding}. Thus, the presence of multitudes of potential localization sites in the stochastic microstructure (\textit{i.e.}, the meso-scale at order of $1 \, mm$) of porous energetic materials needs to be simulated to model SDT \citep{bernecker1986deflagration}. 

\begin{figure}
     \centering
     \begin{subfigure}{\textwidth}
         \centering
         \includegraphics[width=\textwidth]{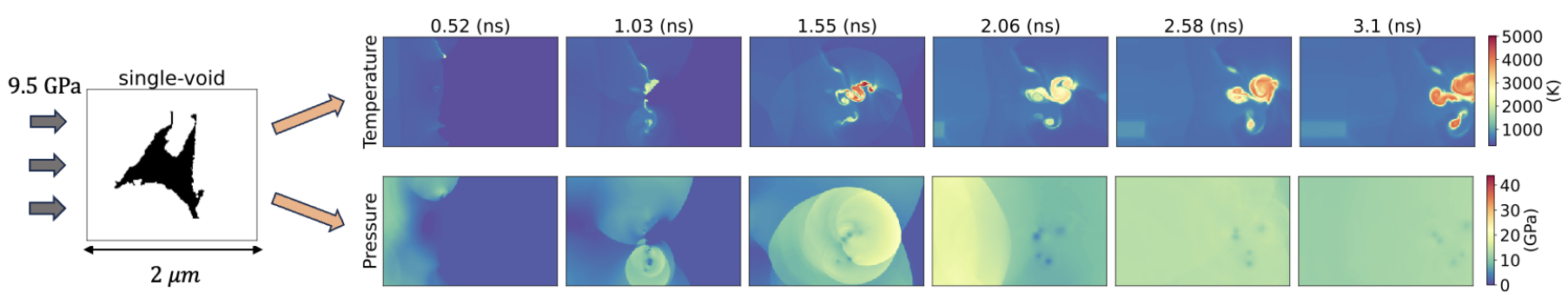}
         \caption{}     
     \end{subfigure}
     \begin{subfigure}{\textwidth}
         \centering
         \includegraphics[width=\textwidth]{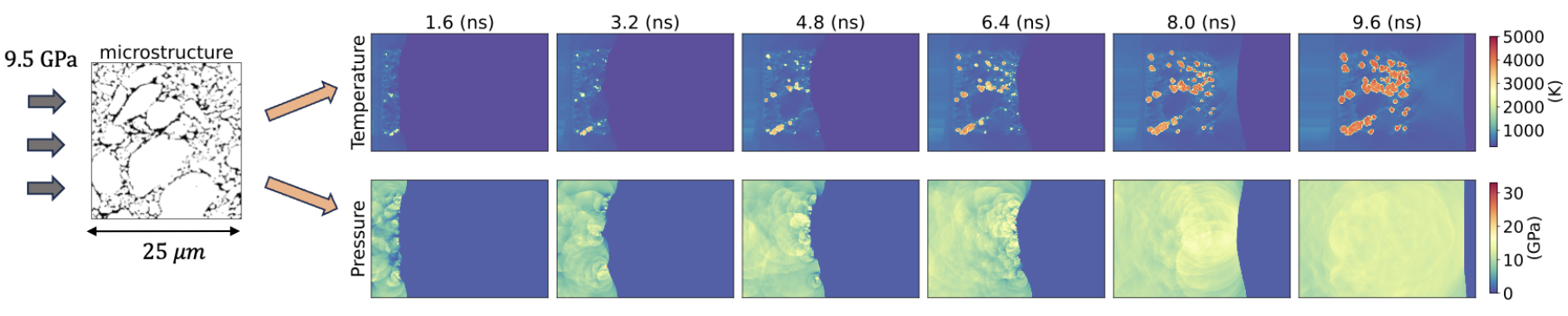}
         \caption{}
     \end{subfigure}
     \caption{Temperature and pressure field evolution corresponding to (a) micro- and (b) meso-scale simulation of shock-induced energy localization in porous energetic material.} 
     \label{fig:sample_T_P_mu}
\end{figure}

Direct numerical simulations (DNS) play an important role in understanding the above-mentioned complex physics at the meso-scale, where defects (pores, cracks) and interfaces in the heterogenegous microstructure play key roles in energy localization \citep{kapahi2013dynamics,rai2015mesoscale,sen2018multi,seshadri2022meso,Nguyen2022multiscale}. 
Figure \ref{fig:sample_T_P_mu} (a) and (b) illustrate an instance of numerical simulation results representing the shock-induced temporal evolution of temperature and pressure fields associated with a micro-scale single-pore collapse \citep{nguyen2022multi} and a meso-scale field of voids in a domain of EM microstructure \citep{seshadri2022meso}, respectively. The columns in the figures show a snapshot of the corresponding temperature and pressure fields at different instants of time (noted above the panels). However, with current computational resources, fully resolved DNS simulations are not feasible for prediction and design in large, \textit{i.e.}, macro-scale, systems that employ energetic materials. 

To enable design and predictive modeling, there has been increasing interest in multi-scale modeling with data-driven approaches, to make fast predictions of the macro-scale dynamics with comparable accuracy to DNS \citep{nguyen2023parc}. Scale bridging between micro- and macro-scales has been accomplished in a variety of ways \citep{fish2010multiscale,liu2006bridging,saurel2018multiscale}; in recent times, data-driven closure models have been developed using highly resolved direct numerical simulations (DNS) to inform machine learning algorithms \citep{sanderse2024scientific,beck2019deep,san2018neural}. The machine learned models serve as surrogates to close macro-scale dynamics. With the exponential rise of deep learning (DL) \citep{lecun2015deep,bengio2017deep}, artificial intelligence (AI)/machine learning (ML) has received a great deal of research attention in the context of scale-bridging for multi-scale, multiphysics problems. In particular, physics-informed machine Learning (PIML) \citep{karniadakis2021physics} approaches seek to assimilate sub-grid dynamics in deep neural networks (DNNs), and a wide range of capabilities for multi-scale modeling using these techniques are now available \citep{nguyen2023parc,kashinath2021physics,oommen2024rethinking}. While DNN-based approaches are powerful for data assimilation and representation, they are inherently data hungry, which in turn places the onus on computationally intensive simulations to supply adequately rich datasets to the AI/ML model. In this regard, the high computational cost of DNS for generating meso-scale numerical simulations continues to pose practical challenges in data-driven approaches.

 Alleviating this need for computationally intensive meso-scale simulations motivates the current paper. Here, we propose a \emph{meta-learning} framework in which one leverages an ensemble of inexpensive micro-scale (inexpensive single pore) simulations in order to accelerate the learning process of a surrogate model for the meso-scale (microstructure-level) dynamics. Broadly speaking, the cross-disciplinary areas of meta-learning \citep{schmidhuber1987evolutionary,bengio1990learning,hospedales2021meta,lee2022meta} and transfer-learning \citep{thrun1998learning,pan2009survey,olivas2009handbook} consist of a class of machine learning methodologies that aim to learn a common ``meta'' information from a family of closely related tasks and use it to quickly adapt the learned model to new unseen tasks. These methods are  based on a \emph{bi-level} learning process in which an \emph{inner (base)} learning algorithm solves a particular task defined by a dataset and objective function, while an \emph{outer (meta)} algorithm updates the inner learning algorithm such that the learned model improves an outer objective/physical solution. In this regard, we use an analogy in natural language processing (NLP) in the following way: an important step in the learning process of an NLP model is \emph{tokenization}, in which a piece of text, \textit{e.g.}, a sentence or a paragraph, is broken down into meaningful chunks of information, \textit{e.g.}, words, which are called \emph{tokens} and can be considered as discrete building blocks of the original text. Tokenization lets an NLP model start by analyzing these building blocks, and in later steps, combine the token-level learned information to capture the correlation between tokens and learn higher-level information about larger segments of text, such as sentences and paragraphs. 

Following this tokenization analogy, we consider the meso-scale dynamics (``paragraphs'') as a \emph{correlated} system of machine learning building blocks, where each constituent encodes a \emph{reduced (latent) representation} of the micro-scale dynamics (``words''). Therefore, we consider the direct sum of the reduced representations of the micro-scale dynamics (single pore collapse simulations as "words"), after applying an appropriate time-delay, as a ``first order approximation'' of the meso-scale dynamics for a short time span.  However, for longer time periods and over larger spatial extents than the micro-scale, the physical fields over neighboring micro-scale building blocks will couple together (akin to interactions between words for constructing coherent ``paragraphs''), so that the resulting meso-scale dynamics is not purely the direct sum of micro-scale dynamics. The effects due to interaction between neighboring micro-scale building blocks are considered as ``higher order corrections.'' In other words,  we consider a decomposition of the meso-scale dynamics in the following (heuristic) form:
\begin{equation} \label{eq:meso-micro-decomposition}
    \text{meso-dynamics} = \left( \bigoplus_{\text{micro-blocks}} \text{micro-dynamics} \times {\Delta t}_{\text{micro}} \right)
    + \left( \text{correlation terms} \times {\Delta t}_{\text{meso}} \right) \,
\end{equation}
where ${\Delta t}_{\text{micro}}$ and ${\Delta t}_{\text{meso}}$ represent the micro- and meso-scale time-periods, respectively (see Section \ref{sec:correlation_scale_bridging} for a more rigorous discussion). Therefore, the proposed meta-learning framework can be summarized into two steps: in the first step, one learns a reduced representation of the micro-scale dynamics, as the building blocks of a ``first order approximation'' of the meso-scale dynamics, from a relatively large dataset of (inexpensive) micro-scale numerical simulations; in the second step, by transferring the learned reduced representation of the micro-scale dynamics to a meso-scale domain, it suffices to learn the correlation between the temporal evolution of micro-scale building blocks. This second step needs training over a relatively small dataset of (expensive) meso-scale simulations compared to directly learning a surrogate meso-scale model from scratch.

The approach studied in this work combines ideas from the two research fields of meta-learning \citep{hochreiter2001learning,koch2015siamese,andrychowicz2016learning,vinyals2016matching,santoro2016meta,ravi2017optimization,franceschi2018bilevel} and reduced order modeling (ROM) \citep{berkooz1993proper,gorban2005invariant,schmid2010dynamic,carr2012applications,quarteroni2014reduced,lee2020model}. To our knowledge, the application of meta-learning approaches to physics problems has been mostly in the context of physics-informed neural networks (PINNs) \citep{raissi2019physics} to improve their performance over a parametrized family of partial differential equations \citep{liu2022novel,psaros2022meta,bihlo2024improving}. In addition, the proposed model architecture in this work, for learning a reduced representation of the micro-scale dynamics, is based on a \emph{probabilistic} formulation of the latent dynamics in contrast to the most of deep learning-based ROMs which consider a \emph{deterministic} latent dynamics \citep{maulik2021reduced,kim2022fast,fries2022lasdi,vlachas2022multiscale,conti2023reduced,bonneville2024gplasdi}. This probabilistic formulation leads to better generalizability and robustness of our model in scarce meso-scale data regime in addition to facilitating uncertainty quantification (UQ) in the prediction results.

The structure of the paper is as follows. In Section \ref{sec:method}, we present the detailed mathematical formulation and architecture of the model considered in this work. In Section \ref{sec:results}, we present our proposed meta-learning framework applied to the problem of learning a surrogate model for the meso-scale dynamics associated with shock-induced energy localization in EM. Furthermore, we compare the prediction performance of our model with a physics-aware recurrent convolutional neural network (PARC), in the scarce meso-scale data regime, in terms of sensitivity quantities of interest (QoI) for EM. Also, the uncertainty in our model's predicted QoI, due the stochastic nature of learned dynamics over the latent space, is investigated. We conclude in Section \ref{sec:discussion} with a summary of our work and potential future directions for expanding the current work. Additionally, we provide details on implementation of our model and its training in Appendix \ref{sec:appendix_implementation}. In Appendix \ref{sec:appendix_DNS}, we discuss more details on the numerical simulation setup considered for generating the data used in this work. Complementary results on the predicted evolution of the temperature $T$ and pressure $P$ fields by our model and PARC for additional test samples are presented in Appendix \ref{sec:appendix_pred_test}.

\section{Method} \label{sec:method}

Consider partial differential equations (PDEs) of the form
\begin{align} \label{PDE-general}
\begin{split}
    &\partial_t \mathbf{u} = \mathcal{F} \left( {x, \mathbf{u}, \partial_x \mathbf{u} , \partial_{xx} \mathbf{u} , \cdots ; \mathfrak{a}} \right) \, , \qquad t \in [0, \mathbb{T}] \, , \; \; x \in \mathbb{X} \, , \\
    &\mathbf{u}(0,x) = \mathbf{u}^0 (x) \, , \qquad x \in \mathbb{X} \, , \\
    &B[\mathbf{u}](t,x) = 0 \, , \qquad t \in [0, \mathbb{T}] \, , \; \; x \in \partial \mathbb{X} \, ,
\end{split}
\end{align}
parametrized by a family of static parameters $\mathfrak{a}$. The nonlinear functional $\mathcal{F}$ in \eqref{PDE-general} encapsulates the governing dynamics of a system of coupled physical fields $\mathbf{u} = (u_1, u_2, \cdots, u_k) : [0, \mathbb{T}] \times \mathbb{X} \to \mathbb{R}^k$ over the spatial domain $\mathbb{X}$ in time interval $[0, \mathbb{T}]$. The notations $\partial_x \mathbf{u}$ and $\partial_{xx} \mathbf{u}$ denote the first- $\frac{\partial \mathbf{u}}{\partial x}$ and second-order $\frac{\partial^2 \mathbf{u}}{\partial x^2}$ partial derivatives w.r.t. the spatial dimension. In addition, $\mathbf{u}^0 (x)$ specifies the initial condition and $B[\mathbf{u}](t,x) = 0$ gives the boundary conditions over the boundary $\partial \mathbb{X}$ of the spatial domain $\mathbb{X}$. The parameters $\psi = (\mathfrak{a}, \partial \mathbb{X}) \in \Psi$ represent physical and geometrical properties of the system, \textit{e.g.}, the material properties and the global shape of the boundary of the spatial domain. 

By discretizing the spatial domain $\mathbb{X}$ into a grid or mesh, the system of PDEs \eqref{PDE-general} leads to a parametrized dynamical system of the following form 
\begin{equation} \label{dynamical-system}
    \frac{d}{dt} \mathbf{U}  = \widetilde{\mathbf{F}} (\mathbf{U}; \psi) \, , \qquad t \in [0, \mathbb{T}] \, , \quad \psi \in \Psi \, ,    
\end{equation}
where $\mathbf{U} = (U_1, U_2, \cdots, U_k) : [0, \mathbb{T}] \to \mathbb{R}^ {k \times N}$ denotes the corresponding high-dimensional \emph{state} vector. This work employs a uniform Cartesian grid over a 2D domain $\mathbb{X} \subset \mathbb{R}^2$. Note that the direct numerical simulations at the micro- and meso-scales are also performed on a uniform fixed Cartesian grid \citep{nguyen2022multi,seshadri2022meso}, and the data provided to the deep learning algorithms are arranged on this grid. This provides a natural connection between the evolution of the fields computed from DNS and the tesselation required for training convolutional neural networks, \textit{i.e.}, the building blocks of the model considered in this work. 

Given a dataset of consecutive time-snapshots 
\begin{equation}
    \left[ \mathbf{U}^{(t_0)} \, , \, \mathbf{U}^{(t_1)} \, , \, \cdots \, , \, \mathbf{U}^{(t_n)} \right] 
    \, , \quad t_{j+1} - t_j = \Delta t \, ,
\end{equation}
of the state vector $\mathbf{U}$ corresponding to various initial conditions and parameter values, we want to learn a surrogate model, which can emulate the time evolution of the dynamical system \eqref{dynamical-system}. We assume that the dominant part of the dynamics of the system happens over a lower-dimensional \emph{invariant manifold (slow manifold)} embedded inside the high-dimensional input space $\mathbb{R}^{k \times N}$ \citep{temam2012infinite, holmes2012turbulence}. Recall that slow manifold \citep{gorban2005invariant,carr2012applications} of a dynamical system is defined as an attracting submanifold of the system's full state space on which the dynamics is slow and free of high-frequency motions compared to the state space regions away from its basins of attraction (see, \textit{e.g.}, \citep{benner2015survey,fresca2021comprehensive,geelen2023operator}).  Hence, we focus on learning a rich enough latent representation of the dynamical system and the corresponding evolution over the latent space. 

\subsection{Architecture Design} \label{sec:architecture_design}

Our probabilistic deep learning-based model consists of two main classes of building blocks, which, roughly speaking, are responsible for 
\begin{enumerate}[\itshape(i)]
    \item learning a nonlinear representation of the slow manifold of the system together with the most probable fluctuations of the dynamics in the neighborhood of the slow manifold; and
    \item learning a stochastic reduced dynamics over the latent space, which captures the essence of the original dynamics in the full state space,
\end{enumerate} 
respectively. In order to achieve the learning task \textit{(i)} in our model, we use \emph{variational autoencoders} (VAE) \citep{kingma2014auto,rezende2014stochastic} for the compression of the high-fidelity fields $U_i$ into corresponding latent fields $Z_i$ which result in better generalizability and robustness of the model in sparse data regimes of the parameter space $\Psi$ compared to the classical autoencoders. For each $i = 1, \cdots , k$, we consider the latent field $Z_i$ as a \emph{discrete Gaussian random field} \citep{rue2005gaussian} whose two-point auto-correlation function over the spatial domain equals the identity matrix, \textit{i.e.}, $Z_i (x_1)$ and $Z_i (x_2)$ are independent multivariate Gaussian random variables for all $x_1 \neq x_2 \in \mathbb{X}$. Hence, the latent field $Z_i$ is uniquely determined by the two fields $\Bar{Z}_i$ and $Z^\sigma_i$, where $\Bar{Z}_i (x)$ and $Z^\sigma_i (x)$ specify the mean and the log-variance of the Gaussian random variable $Z_i (x)$, $\forall x \in \mathbb{X}$. We consider the space of pairs ${\left( \Bar{Z}_i , Z^\sigma_i \right)}$ of the above-mentioned form as the latent space associated with each field $U_i$, $i = 1, \cdots , k$. In other words, the latent mean fields $\Bar{Z}_i$ can be considered as elements of the slow manifold of the corresponding dynamical system, while the latent log-variance fields $Z^\sigma_i$ capture the fluctuations of the dynamics near the slow manifold.  

The encoder ${E_i = \left( {\bar{E}_i , E^\sigma_i } \right) : U_i \mapsto {\left( \Bar{Z}_i , Z^\sigma_i \right)}}$ of our VAE is considered to be a convolutional neural network (CNN) which encode the high-fidelity field $U_i$ into the pair $(\Bar{Z}_i, Z^\sigma_i)$ in the following way:
\begin{equation} \label{mean-field}
    \Bar{Z}_i (x) = \left( {\Bar{K}_{\theta_i} \ast U_i} \right) (x)
                = \int_{\mathbb{X}} {\Bar{K}_{\theta_i} (x-y) \cdot U_i(y)} \, \mathrm{d} y 
                \, , \quad x \in \mathbb{X} \, ,
\end{equation}
and 
\begin{equation} \label{var-field}
    Z^\sigma_i (x) = \left( {{K^\sigma_{\theta'_i}} \ast U_i} \right) (x)
                = \int_{\mathbb{X}} {{K^\sigma_{\theta'_i}} (x-y) \cdot U_i(y)} \, \mathrm{d} y 
                \, , \quad x \in \mathbb{X} \, .
\end{equation}
Let $d$ be the dimension of each latent field $Z_i$, $i=1, \cdots , k$ over the spatial domain $\mathbb{X}$, \textit{i.e.}, the number of channels of the corresponding numerical tensor. In general, the convolution kernels $\Bar{K}_{\theta_i}$ and ${K^\sigma_{\theta'_i}}$, in \eqref{mean-field} and \eqref{var-field}, can be considered as $d$-by-$1$ matrix-valued functions over $\mathbb{X} \times \mathbb{X}$, \textit{i.e.}, 
\begin{equation} \label{eq:encoder_convolution_1}
    \Bar{K}_{\theta_i} : \mathbb{X} \times \mathbb{X} \to \mathcal{M}_{d \times 1} (\mathbb{R})
\end{equation}
and 
\begin{equation} \label{eq:encoder_convolution_2}
    {K^\sigma_{\theta'_i}} : \mathbb{X} \times \mathbb{X} \to \mathcal{M}_{d \times 1} (\mathbb{R}) \, ,
\end{equation}
respectively, where $\mathcal{M}_{d \times 1} (\mathbb{R})$ denotes the set of all $d$-by-$1$ real matrices. In the particular setup of convolutional neural networks, one considers the convolution kernels to be translation-invariant matrix-valued functions with finite support, \textit{i.e.}, a fixed-size sliding window which sweeps the spatial domain $\mathbb{X}$. The entries of the corresponding matrices form the set of trainable parameters $(\theta_i , \theta'_i)$ of the corresponding models which is learned through the training process. The latent mean field $\bar{Z}_i (x)$ (or, similarly, the log-variance field $Z^\sigma_i (x)$) at each $x \in \mathbb{X}$ is a weighted sum of $d$-dimensional vectors ${\Bar{K}_{\theta_i} (x,y) \cdot U_i(y)}$ which are result of action of the matrix $\Bar{K}_{\theta_i} (x,y)$ on the $1$-dimensional vector $U_i (y)$, $\forall y \in \mathbb{X}$.

In addition, the decoder ${D_i : Z_i \mapsto \widehat{U}_i}$ of our VAE is also considered to be a convolutional neural network which generate the reconstructed field $\widehat{U}_i$ in the input space from the latent field $Z_i$ as follows: 
\begin{equation} \label{recon-field}
    \widehat{U}_i (x) = \left( {{K_{\varphi_i}} \ast Z_i} \right) (x)
                = \int_{\mathbb{X}} {{K_{\varphi_i}} (x-y) \cdot Z_i(y)} \, \mathrm{d} y 
                \, , \quad x \in \mathbb{X} \, .
\end{equation}

In the next step, in order to achieve the learning task \textit{(ii)}, the reduced dynamics (a.k.a. latent evolution) is learned through an \emph{autoregressive} model
\begin{equation} \label{autoregression-1}
    {\left( {\bar{\mathbf{Z}} , {\mathbf{Z}}^{\sigma}} \right)}_{|t_{j+1}} =
    \mathbf{F}_\alpha \left(
    {\left( {\bar{\mathbf{Z}} , {\mathbf{Z}}^{\sigma}} \right)}_{|t_{j}} \, ; \psi
    \right) \, , \quad \psi \in \Psi \, ,
\end{equation}
where $\bar{\mathbf{Z}} = {\left( \bar{Z}_1 , \bar{Z}_2 , \cdots , \bar{Z}_k \right)}$, ${\mathbf{Z}}^\sigma = {\left( {Z^\sigma_1 , Z^\sigma_2 , \cdots , Z^\sigma_k} \right)}$, and $\alpha$ denotes the trainable parameters of the model. In the following, we assume that all the corresponding autoregressive functionals depend on the fixed parameters $\psi \in \Psi$ together with a set of trainable parameters $\alpha$, and do not mention this dependence explicitly to avoid cumbersome notation. Motivated by the Stochastic Differential Equation (SDE) framework, we decompose the dynamics induced by the mapping $\mathbf{F}$, in \eqref{autoregression-1}, into a \emph{deterministic} part
\begin{equation} \label{deterministic-dynamics-1}
    {\bar{\mathbf{Z}}}_{|t_{j+1}} = \bar{\mathbf{F}} \left( {\bar{\mathbf{Z}}}_{|t_{j}} \right) \, ,
\end{equation}
and a \emph{stochastic} part
\begin{equation} \label{stochastic-dynamics-1}
    {{\mathbf{Z}}^\sigma_{|t_{j+1}}} = \mathbf{F}^\sigma \left(
    {\left( {\bar{\mathbf{Z}} , {\mathbf{Z}}^{\sigma}} \right)}_{|t_{j}} 
    \right) \, .
\end{equation}

Furthermore, we decompose the mapping $\bar{\mathbf{F}}$ representing the deterministic part of the latent evolution, \textit{i.e.}, \eqref{deterministic-dynamics-1}, in the following form
\begin{equation} \label{deterministic-dynamics-2}
    \bar{\mathbf{F}} \left( {\bar{\mathbf{Z}}} \right) 
    = {\bar{F}}_{\mathrm{corr}}  \left({{\bar{F}}_{\mathrm{dec}} \left( {\bar{\mathbf{Z}}} \right)}\right) \, ,
\end{equation}
in which
\begin{equation} \label{deterministic-dynamics-3}
    {\bar{F}}_{\mathrm{dec}} \left( {\bar{\mathbf{Z}}} \right) = 
    \begin{bmatrix}
        \bar{f}_1 & & &  \\[0.25em]
         & \bar{f}_2 & & \\[0.25em]
         & & \ddots & \\[0.25em]
         & & &  \bar{f}_k
    \end{bmatrix}
    \begin{bmatrix}
        \bar{Z}_1 \\[0.25em] \bar{Z}_2 \\[0.25em] \vdots \\[0.25em] \bar{Z}_k
    \end{bmatrix} 
\end{equation}
captures the \emph{decoupled} evolution of each latent mean-field $\bar{Z}_i$, $i=1, \cdots , k$, independently, while ${\bar{F}}_{\mathrm{corr}}$ encapsulates the \emph{correlation} (interaction) among all the latent mean-fields. Similarly, the mapping $\mathbf{F}^\sigma$ representing the stochastic part of the latent evolution, \textit{i.e.}, \eqref{stochastic-dynamics-1}, is decomposed into three building blocks as the following
\begin{equation} \label{stochastic-dynamics-2}
    \mathbf{F}^\sigma {\left( {\bar{\mathbf{Z}} , {\mathbf{Z}}^{\sigma}} \right)} 
    = {F^\sigma_{\mathrm{corr}}} \left(
    {\bar{F}}_{\mathrm{dec}} \left( {\bar{\mathbf{Z}}} \right) \, , \,
    {F^\sigma_{\mathrm{dec}}} {\left( {\bar{\mathbf{Z}} , {\mathbf{Z}}^{\sigma}} \right)}
    \right) \, ,
\end{equation}
where ${\bar{F}}_{\mathrm{dec}}$ is given by \eqref{deterministic-dynamics-3}, and
\begin{equation} \label{stochastic-dynamics-3}
    {F^\sigma_{\mathrm{dec}}} {\left( {\bar{\mathbf{Z}} , {\mathbf{Z}}^{\sigma}} \right)} =
    \begin{bmatrix}
        f^\sigma_1 & & &  \\[0.28em]
         & f^\sigma_2 & & \\[0.28em]
         & & \ddots & \\[0.28em]
         & & &  f^\sigma_k
    \end{bmatrix}
    \begin{bmatrix}
        (\bar{Z}_1 , Z^\sigma_1) \\[0.25em] (\bar{Z}_2 , Z^\sigma_2) \\[0.25em] \vdots \\[0.25em] (\bar{Z}_k , Z^\sigma_k) 
    \end{bmatrix} \, .
\end{equation}
The mapping ${F^\sigma_{\mathrm{corr}}}$ in \eqref{stochastic-dynamics-2} models the correlated evolution of all the latent log-variance-fields $Z^\sigma_i$, $i=1, \cdots , k$. 

We consider a convolutional neural network \citep{schmidhuber2015deep,mallat2016understanding} (CNN)-based architecture for all the building blocks of our model, \textit{i.e.}, the variational encoder, decoder, and the autoregressive functionals in \eqref{deterministic-dynamics-2} and \eqref{stochastic-dynamics-2}, in order to make our model to be transferable from micro- to meso-scale. For the results presented in this paper, we consider a VGG-Net \citep{simonyan2014very} and a U-Net \citep{ronneberger2015u} architecture for the autoencoder and the latent dynamics autoregressive functionals, respectively. It is a multi-level, multi-resolution architecture which captures the local correlation between micro-scale building blocks in a hierarchical manner. Note that the choice of CNN-based architecture is due to its \emph{local connectivity} which leads to learn spatially local patterns in the input fields, and its ability to learn non-local effects as the depth of the architecture increases. Alternative well-established architectures for learning local and non-local patterns  are Fourier Neural Operator (FNO) \citep{li2020fourier} and Vision Transformer (ViT) \citep{dosovitskiy2020image}. However, these two architectures are computationally more expensive and less data efficient compared to CNNs which makes their applicability to the present scarce data scenario to be challenging. Details of the implementation of our CNN-based model are discussed in Appendix \ref{sec:appendix_implementation}. Source codes are available online at \href{https://github.com/ShahabAzarfar/Multiscale-Latent-Dynamics}{https://github.com/ShahabAzarfar/Multiscale-Latent-Dynamics}.

\subsection{Training}

For training each VAE in our model, the corresponding trainable parameters $(\theta_i , \theta'_i , \varphi_i)$ of the convolution kernels ${\Bar{K}_{\theta_i}}$, ${K^\sigma_{\theta'_i}}$ and ${K_{\varphi_i}}$, in \eqref{mean-field}, \eqref{var-field} and \eqref{recon-field}, respectively, are learned from the data by minimizing the reconstruction loss
\begin{equation} \label{loss-recon}
    \mathcal{L}_{\mathrm{recon}} = {\left\lVert {U_i - \widehat{U}_i} \right\rVert}_{L^2}^2 
    = \int_{\mathbb{X}} {\left| {U_i (x) - \widehat{U}_i (x)} \right|}^2 \, \mathrm{d}x
\end{equation}
together with the Kullback-Leibler divergence loss\footnote{The $c$-th component of the vectors $\Bar{Z}_i (x)$ and $Z^\sigma_i (x)$ are denoted by $\Bar{Z}_{i,c} (x)$ and $Z^{\sigma}_{i,c} (x)$, respectively.} 
\begin{equation} \label{loss-KL}
    \mathcal{L}_{\mathrm{KL}} = \frac{1}{2} \int_{\mathbb{X}} \sum_c 
    \left[  {\left( \Bar{Z}_{i,c} (x) \right)}^2 + \exp{\left( Z^{\sigma}_{i,c} (x) \right)} - Z^{\sigma}_{i,c} (x) -1  \right] 
    \, \mathrm{d} x \, .
\end{equation}
Note that since we are considering each latent field $Z_i$ to be a discrete Gaussian random field with trivial two-point auto-correlation function, \textit{i.e.}, equal to identity matrix, the Kullback-Leibler divergence loss \eqref{loss-KL} is essentially the Kullback-Leibler divergence between a direct sum of univariate Gaussian distributions and the standard normal distributions. We train the VAE associated with each field $U_i$, $i=1, \cdots , k$, \emph{independently}, and use the trained encoder to compress $U_i$ into the corresponding latent representation ${\left( \Bar{Z}_i , Z^\sigma_i \right)}$.

Let ${\mathbf{E} = \left( {\bar{\mathbf{E}} , {\mathbf{E}}^\sigma} \right): \mathbf{U} \mapsto (\bar{\mathbf{Z}} , {\mathbf{Z}}^\sigma)}$ and ${\mathbf{D}: \mathbf{Z} \mapsto \widehat{\mathbf{U}}}$ be the direct sum of the VAE-encoders and the VAE-decoders, respectively, corresponding to the fields $U_i$, $i=1, \cdots , k$. Considering \eqref{autoregression-1}, the autoregressive rollout of our model for $M$ time-steps can be written as
\begin{equation} \label{autoregressive-rollout}
    {\widetilde{\mathbf{U}}}^{(t_{j+M})} = 
    \mathbf{D} \left(
    {\mathbf{F}_\alpha (\cdot ; p)}^{(M)}
    \left( \mathbf{E} \left({{\mathbf{U}}^{(t_j)}}\right) \right)
    \right) 
    =
    \mathbf{D} \left(
    \underbrace{{\mathbf{F}_\alpha (\cdot ; p)} \circ \cdots \circ {\mathbf{F}_\alpha (\cdot ; p)}}_{\text{composing $M$ times}} \left( \mathbf{E} \left({{\mathbf{U}}^{(t_j)}}\right) \right)
    \right) \, .
\end{equation}
As the learning objective for training of the autoregressive model ${\mathbf{F}_\alpha (\cdot ; p)}$, we consider the following three loss functions which are extensions of the loss functions introduced in \citep{wu2022learning}:
\begin{equation} \label{loss-multi-step}
    \mathcal{L}_{\mathrm{multi-step}} = \sum_{m=1}^M w_m \, \ell \left( {\mathbf{U}}^{(t_{j+m})} , {\widetilde{\mathbf{U}}}^{(t_{j+m})} \right) \, ,
\end{equation}
\begin{equation} \label{loss-consistency-mean}
    \mathcal{L}_{\mathrm{consistency-mean}} = \sum_{m=1}^M 
    \frac{ {\left\lVert { {\bar{\mathbf{F}}}^{(m)} \circ 
    \bar{\mathbf{E}}  \left( {\mathbf{U}}^{(t_{j})} \right) 
    - \bar{\mathbf{E}}  \left( {\mathbf{U}}^{(t_{j+m})} \right)} \right\rVert}_{L^2}^2 }
    { {\left\lVert {\bar{\mathbf{E}}  \left( {\mathbf{U}}^{(t_{j+m})} \right)} \right\rVert}_{L^2}^2 } \, ,
\end{equation}
and
\begin{equation} \label{loss-consistency-var}
    \mathcal{L}_{\mathrm{consistency-var}} = \sum_{m=1}^M 
    \frac{ {\left\lVert { {{\mathbf{F}}^\sigma}^{(m)} \circ 
    {\mathbf{E}}  \left( {\mathbf{U}}^{(t_{j})} \right) 
    - {{\mathbf{E}}^\sigma}  \left( {\mathbf{U}}^{(t_{j+m})} \right)} \right\rVert}_{L^2}^2 }
    { {\left\lVert {{{\mathbf{E}}^\sigma}  \left( {\mathbf{U}}^{(t_{j+m})} \right)} \right\rVert}_{L^2}^2 } \, .
\end{equation}
In \eqref{loss-multi-step}, we perform $m$-step latent evolution and compare the reconstructed fields ${\widetilde{\mathbf{U}}}^{(t_{j+m})}$, given by \eqref{autoregressive-rollout}, with the target fields ${\mathbf{U}}^{(t_{j+m})}$ in the \emph{input} space, up to time horizon $M$. Here, the loss function $\ell$ can typically be MSE, MAE, or $L_2$ loss, and $w_m$ denotes the weights for each time-step. In \eqref{loss-consistency-mean}, we compare the $m$-step latent rollout of the encoded latent mean-fields 
${ {\bar{\mathbf{F}}}^{(m)} \circ \bar{\mathbf{E}}  \left( {\mathbf{U}}^{(t_{j})} \right)}$ with the target latent mean-field ${\bar{\mathbf{E}}  \left( {\mathbf{U}}^{(t_{j+m})} \right)}$ in the \emph{latent} space. The loss function \eqref{loss-consistency-var} is similar to \eqref{loss-consistency-mean} except the same process is applied to the latent log-variance-fields ${{\mathbf{Z}}^\sigma}$ instead of the latent mean-fields $\bar{\mathbf{Z}}$. 
    
\section{Results} \label{sec:results}

The thermomechanics governing the evolution of physical fields over a meso-scale domain (\textit{i.e.}, size of  $O (100 \, \mu m)$) of shock-loaded porous EM is the result of complex interactions between micro-scale dynamics resulting from the collapse of individual pores \citep{saurel1999multiphase}. The full set of governing equations for the corresponding dynamics is given as a system of hyperbolic PDEs, which involves the conservation laws for mass, momentum, and energy together with appropriate constitutive models such as the equation of state and flow rules. It describes the temporal evolution of a set of coupled physical fields, including the density, temperature, and pressure scalar fields; the velocity vector field; and the stress and strain 2-tensor fields (see \citep{sambasivan2013simulation,kapahi2013dynamics,rai2017high} for details). The direct numerical simulation (DNS) dataset used in this paper is generated by solving the above-mentioned full system of coupled PDEs. More details on the numerical simulation setup are provided in Appendix \ref{sec:appendix_DNS}. In this work, following \citep{nguyen2023parc}, we consider the dynamical system corresponding to the evolution of the temperature $T$, pressure $P$ and microstructural morphology $\mu$ fields which are more directly related to the evolution of hotspots to quantify energy localization at the meso-scale to determine the macro-scale shock sensitivity quantities of interest for EM \citep{udaykumar2020unified}. These three fields play key roles in so-called reactive burn models which are commonly used to drive shock-to-detonation calculations in production hydrocodes \citep{mader2007numerical}. Our goal is to learn a surrogate model for the coupled dynamics of these three fields over a meso-scale domain in EM microstructures. 

We follow a \emph{meta-learning} approach to leverage a relatively large dataset of inexpensive micro-scale simulations, corresponding to single-pore collapse, combined with a small dataset of meso-scale simulations for the above-mentioned learning process. The proposed meta-learning framework is based on the idea that the meso-scale dynamics can be considered as a correlated system of interacting micro-scale building blocks, where the constituents can be learned from the underlying micro-scale physics. In this approach, we start by learning a latent representation of the micro-scale dynamics from a relatively large dataset ($O(100)$ samples) of micro-scale single pore collapse simulations. These machine-learned latent representations are considered as the building blocks of a ``first order approximation'' of the meso-scale dynamics. In the next step, in order to learn the ``higher order corrections'' corresponding to the correlation between the micro-scale building blocks, we train our model over a relatively small ($<10$ samples) dataset of meso-scale simulations. We elaborate on the above-mentioned two steps of our proposed meta-learning framework in Section \ref{sec:micro-dynamics} and \ref{sec:correlation_scale_bridging}, respectively. 

\subsection{Learning a latent representation of micro-scale dynamics} \label{sec:micro-dynamics} 

In this section, we describe the learning process of a reduced representation of the micro-scale dynamics of the coupled fields $T$, $P$ and $\mu$ corresponding to the shock-induced deformation in a 2D-domain of porous energetic material containing an isolated single-void (see Figure \ref{fig:sample_T_P_mu} (a)). We consider a probabilistic deep learning-based model following the architecture discussed in Section \ref{sec:architecture_design}. 

We start by learning a nonlinear representation of the slow manifold of the system. To this end, we train three independent VAEs, associated with $T$, $P$, and $\mu$ fields, on the single-void data in order to construct latent representations of the form $(\bar{Z}_T, Z^\sigma_T)$, $(\bar{Z}_P, Z^\sigma_P)$ and $(\bar{Z}_\mu, Z^\sigma_\mu)$, respectively. The encoder corresponding to each physical field in our model compresses the input field into a latent field with 64 times lower spatial resolution and four channels, which leads to a total compression rate equal to $1/16$. In other words, if we interpret the value of the involved physical fields in each spatial grid (pixel) as a degree of freedom of the corresponding multiphysics system, the autoencoder of our model learns a latent manifold embedded into the input space whose dimension equals $1/16$-th of the input space dimensionality. This level of dimensionality reduction makes it possible for our autoregressive model to learn a reduced representation of the original micro-scale dynamics which captures the main patterns of the underlying dynamics while ignoring its detailed high-frequency part. 

\begin{figure}
    \centering
    \includegraphics[width=0.5\linewidth]{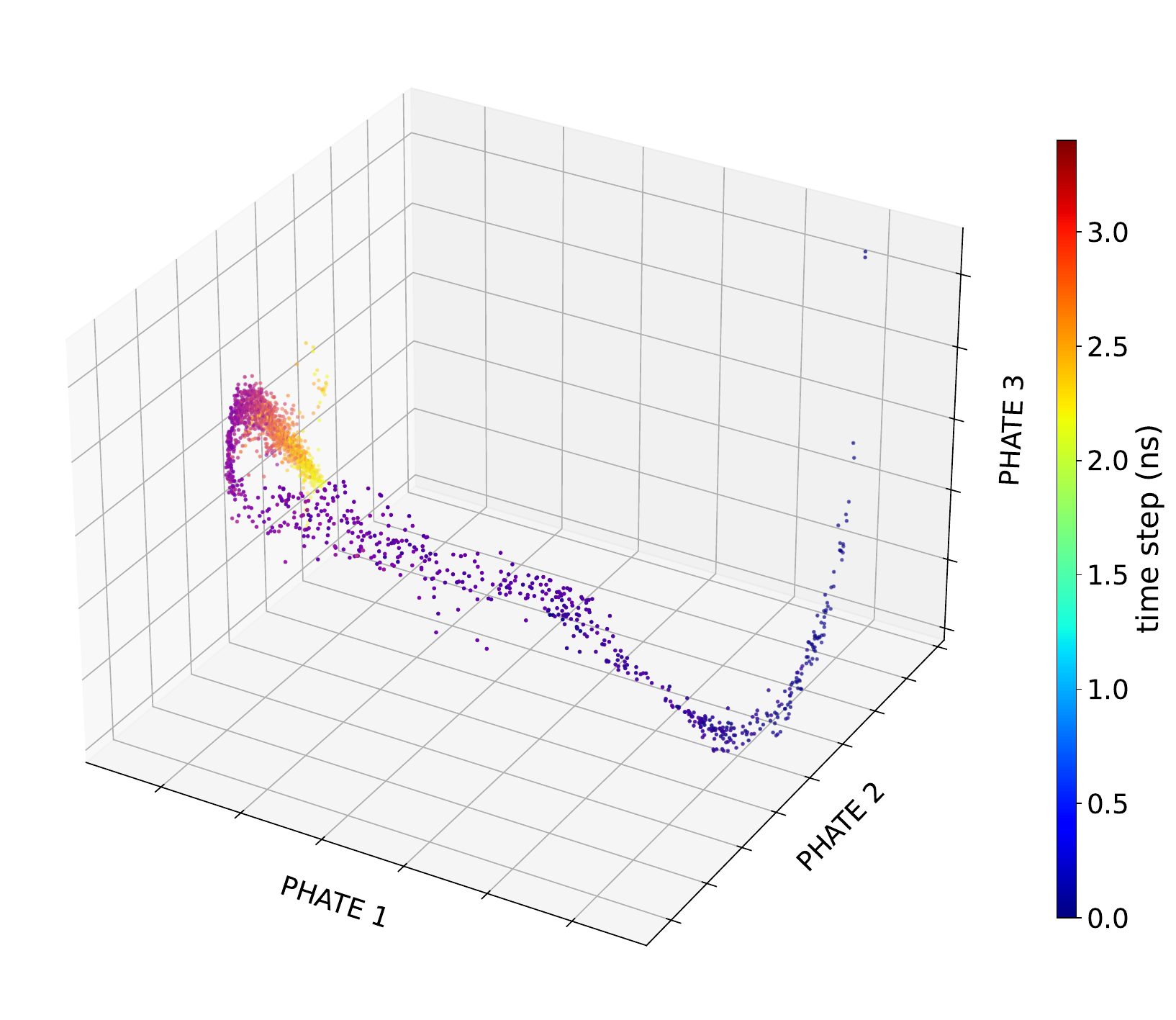}
    \caption{3D visualization of the slow manifold of temperature latent mean fields $\bar{Z}_T (t)$}
    \label{fig:latent_manifold}
\end{figure}

A visual 3D-embedding of the slow manifold of temperature latent mean fields $\bar{Z}_T$, corresponding to the hotspot ignition and growth due to shock-induced single pore collapse in EM, is shown in Figure~\ref{fig:latent_manifold}.\footnote{We used a nonlinear dimensionality reduction technique, called PHATE \citep{moon2019visualizing}, for visualizing the manifold embedding. It is based on the Diffusion Maps algorithm \citep{coifman2006diffusion} which defines an intrinsic notion of distance between the data points.} One can see that the latent mean fields $\bar{Z}_T$ are initially quite close to each other which reflects the fact that, starting from the constant initial condition $T=300 \mathrm{K}$, the temperature fields of all sample simulations, corresponding to the collapse of voids with various sizes and shapes, remain quite similar before the arrival of the shock front at the void which initiates the void-collapse and formation of hotspots. However, in later time-steps, the corresponding slow manifold expands and bifurcates into several branches depending on the temperature distribution and growth of the resulting hotspot.

\begin{figure}
    \centering
    \includegraphics[width=0.8\linewidth]{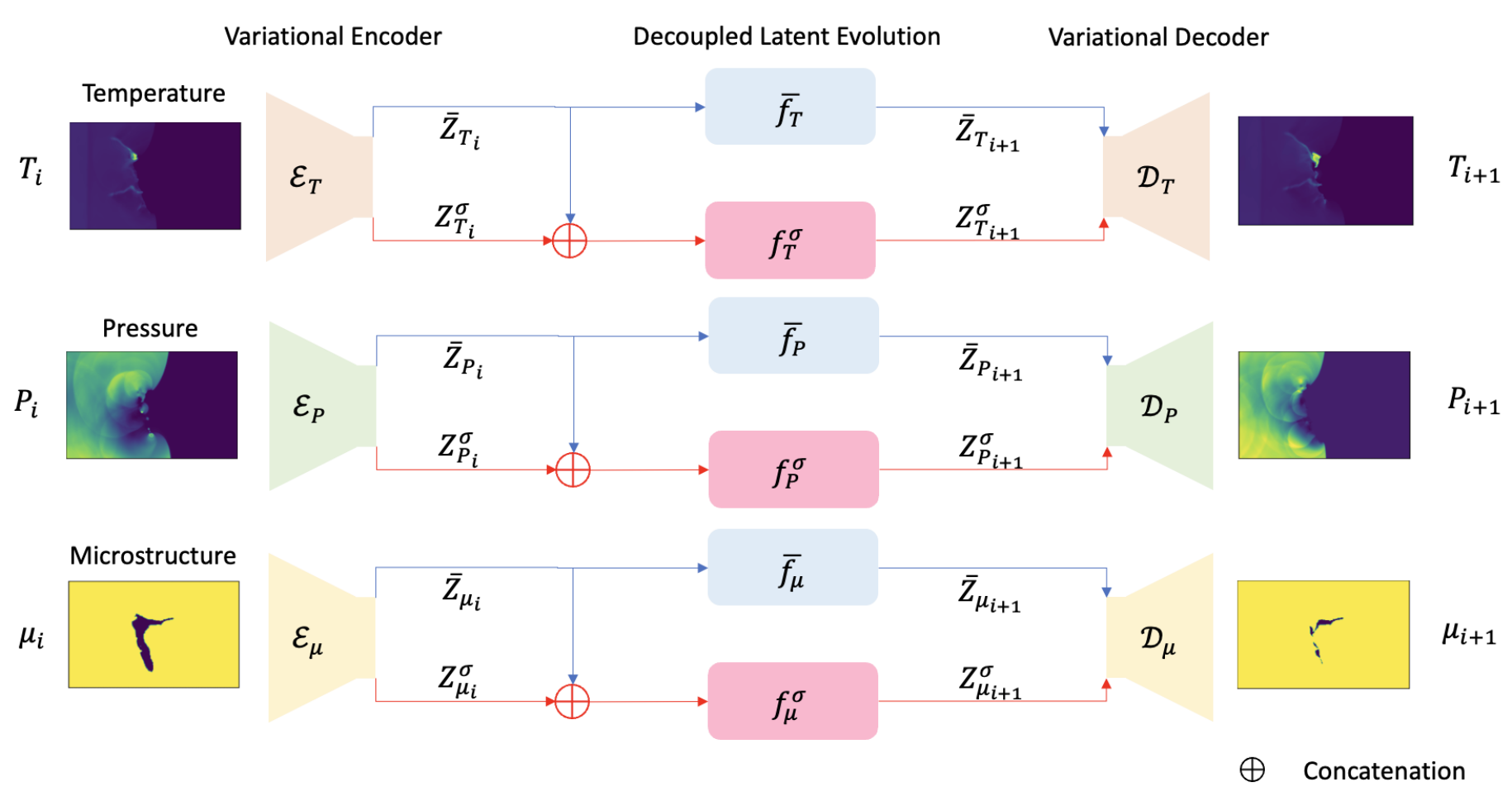}
    \caption{Schematic diagram of the decoupled latent evolution model}
    \label{fig:decoupled model}
\end{figure}

In the next step, we want to learn the stochastic reduced dynamics of the encoded latent fields $(\bar{Z}_U , Z^\sigma_U)$, where $U$ denotes one of the physical fields $T, P , \mu$. Considering \eqref{deterministic-dynamics-2} and \eqref{stochastic-dynamics-2}, it consists of learning the decoupled latent evolution mappings $({\bar{F}_{\mathrm{dec}}}, {F^\sigma_{\mathrm{dec}}})$, given by \eqref{deterministic-dynamics-3} and \eqref{stochastic-dynamics-3}, in addition to the field-correlation mappings $({\bar{F}_{\mathrm{corr}}}, {F^\sigma_{\mathrm{corr}}})$. We start by learning the decoupled evolution of each latent field $(\bar{Z}_U , Z^\sigma_U)$, given by the associated autoregressive functionals $(\bar{f}_U , f^\sigma_U)$, independently, where $U = T, P , \mu$. A schematic diagram of the decoupled latent evolution models, along with the corresponding variational encoder and decoder, is illustrated in Figure \ref{fig:decoupled model}. The loss functions for training of each $(\bar{f}_U , f^\sigma_U)$ are the same loss functions $\mathcal{L}_{\mathrm{multi-step}}$, $\mathcal{L}_{\mathrm{consistency-mean}}$ and $\mathcal{L}_{\mathrm{consistency-var}}$, given by \eqref{loss-multi-step}, \eqref{loss-consistency-mean} and \eqref{loss-consistency-var}, respectively, in the case of a single field $U$. The direct sum of the learned decoupled latent evolution models $(\bar{f}_U , f^\sigma_U)$, for $U = T, P , \mu$, is considered as $({\bar{F}_{\mathrm{dec}}}, {F^\sigma_{\mathrm{dec}}})$. 

We notice that the decoupled latent evolution models can only encapsulate the \emph{local} behavior of the underlying dynamical system corresponding to the evolution of temperature $T$, pressure $P$, and microstructural morphology $\mu$ fields. For example, although the short rollout of the autoregressive mapping $(\bar{f}_T , f^\sigma_T)$ can capture, to some extent, the advection and diffusion of a hotspot after its formation, its long rollout starting from $t=0$ can hardly make any stable prediction of the temperature evolution. Hence, it is of fundamental importance to learn the mutual correlation and interaction between the three evolving physical fields $T$, $P$ and $\mu$ which is captured by the latent evolution mapping $(\bar{F}_{\mathrm{corr}}, F^\sigma_{\mathrm{corr}})$. 

\begin{figure}
    \centering
    \includegraphics[width=0.8\linewidth]{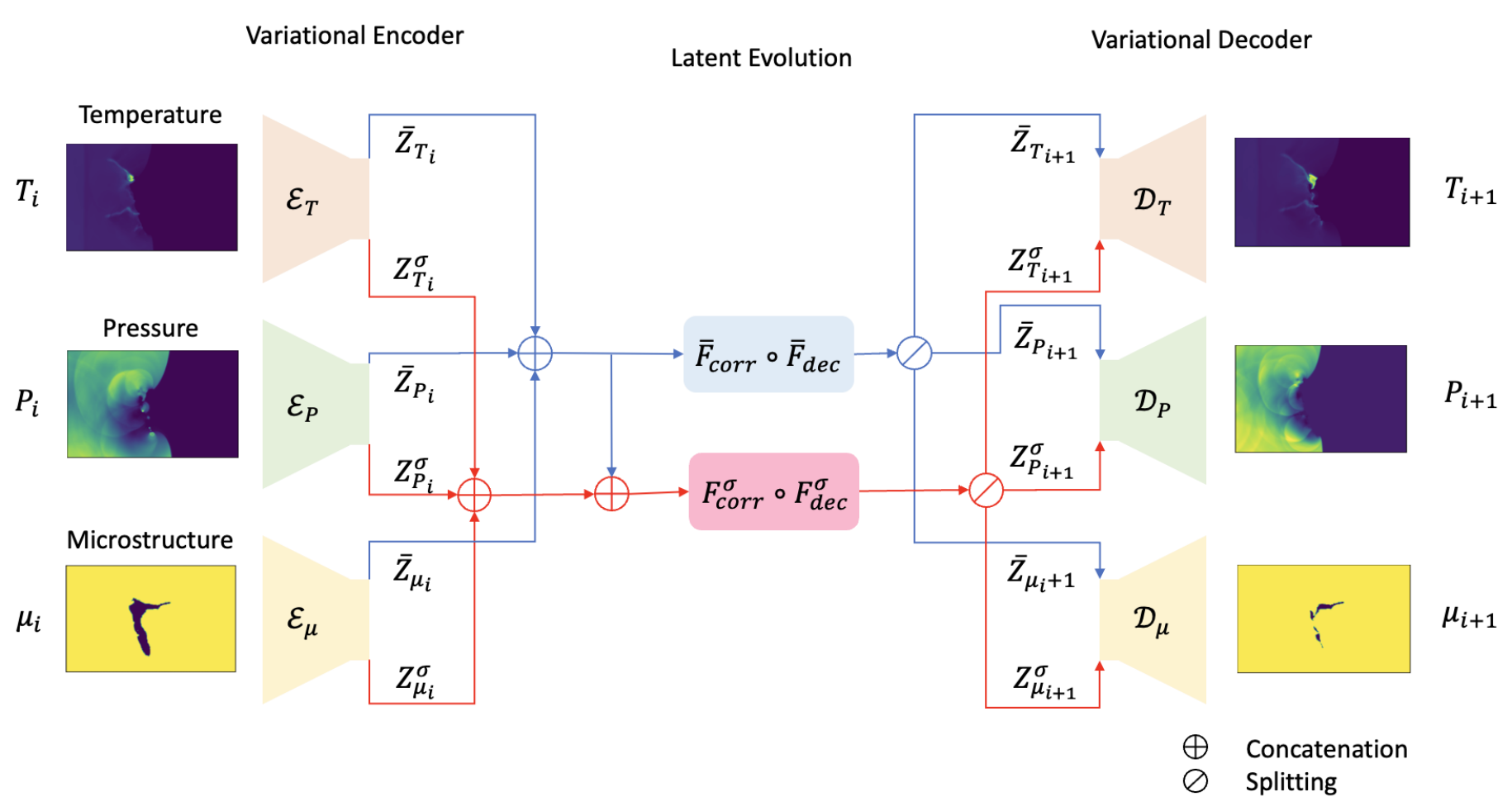}
    \caption{Schematic diagram of the combination of three VAEs with the autoregressive latent evolution mappings in our full model.}
    \label{fig:coupled_model_full}
\end{figure}

We consider the full autoregressive functional as the composition of $(\bar{F}_{\mathrm{corr}}, F^\sigma_{\mathrm{corr}})$ with $({\bar{F}_{\mathrm{dec}}}, {F^\sigma_{\mathrm{dec}}})$ according to \eqref{deterministic-dynamics-2} and \eqref{stochastic-dynamics-2}. The deterministic part of the correlated dynamics of the latent mean fields $\bar{Z}_T$, $\bar{Z}_P$ and $\bar{Z}_\mu$ is learned by the mapping $\bar{F}_{\mathrm{corr}}$. On the other hand, the mapping $F^\sigma_{\mathrm{corr}}$ encapsulates the inherent stochasticity in the coupled evolution of all the corresponding latent fields $(\bar{Z}_U , Z^\sigma_U)$ for $U = T, P , \mu$. A detailed illustration of the way in which various building blocks of the latent autoregressive functionals and the three variational autoencoders are combined to construct our full model is given in Figure \ref{fig:coupled_model_full}. 

The composition of $(\bar{F}_{\mathrm{corr}}, F^\sigma_{\mathrm{corr}})$ with $({\bar{F}_{\mathrm{dec}}}, {F^\sigma_{\mathrm{dec}}})$ makes our model capable of learning the \emph{global}, in addition to the local, dynamics of the underlying coupled physical fields. In other words, starting from the initial condition for the temperature $T$, pressure $P$ and microstructural morphology $\mu$ fields at $t=0$, the long-time rollout of the full autoregressive model can predict the evolution of all three fields with reasonable accuracy. Note that the coupled dynamics of these fields involves, at least, two quite distinct phases \citep{lee1980phenomenological,menikoff2010reactive}: i) hotspot-ignition phase: the time period up to complete collapse of voids which results in ignition and formation of high temperature regions called hotspots; ii) hotspot-growth phase: the time period in which the 
highly intense energy localized in the hotspots start to diffuse though the energetic material. The ``kinks'' in the slow manifold of $\bar{Z}_T$, illustrated in Figure \ref{fig:latent_manifold}, can be interpreted as the transition point between these different phases of the underlying dynamics. The existence of these ``phase transitions'' makes learning the long-term (global) dynamics of the coupled fields $T$, $P$, and $\mu$, a challenging task.  

\begin{figure}
     \centering
     \begin{subfigure}{\textwidth}
         \centering
         \includegraphics[width=\linewidth]{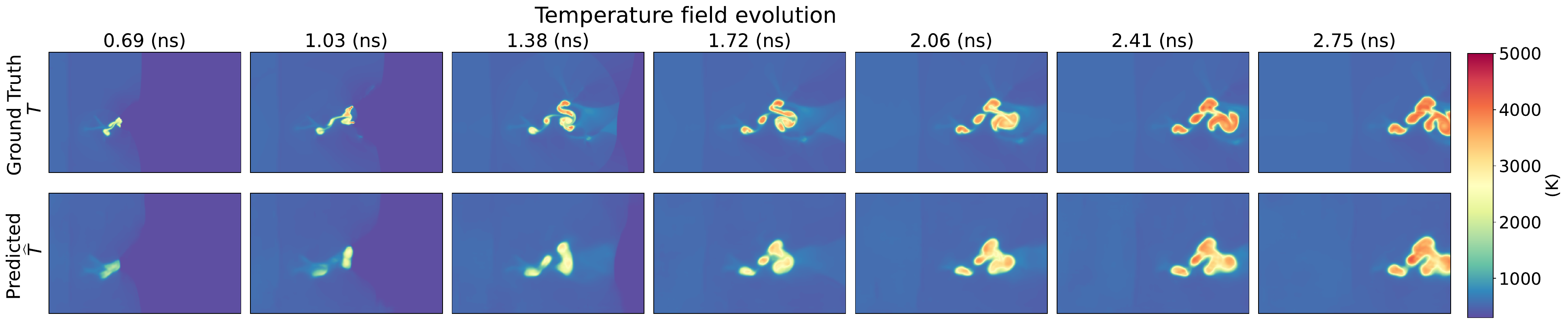}
         \caption{}
     \end{subfigure}
     \begin{subfigure}{\textwidth}
         \centering
         \includegraphics[width=\linewidth]{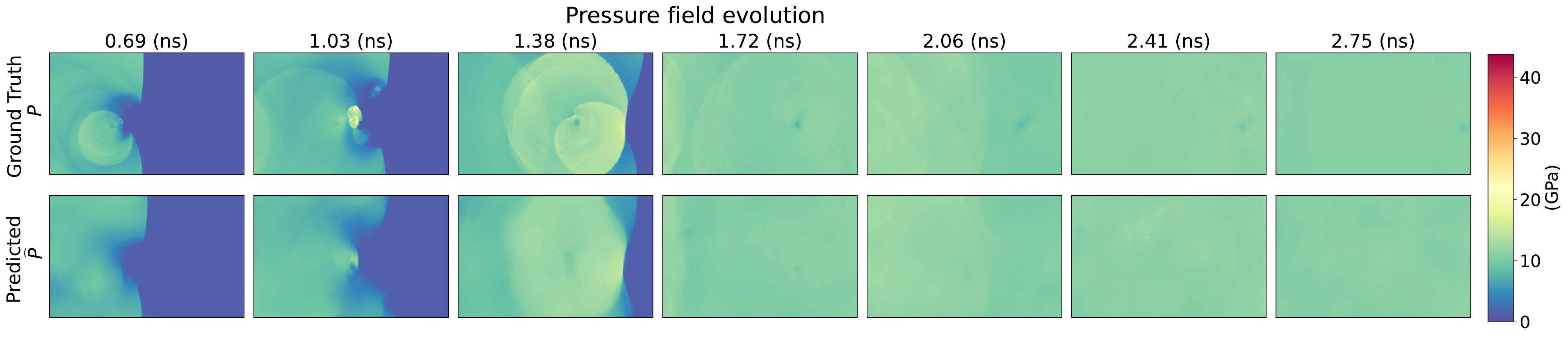}
         \caption{}
     \end{subfigure}
     \caption{Our model's prediction of the (a) temperature and (b) pressure field evolution corresponding to shock-induced single-pore collapse in EM.}
     \label{fig:micro_prediction}
\end{figure}

A direct numerical simulation of the evolution of the temperature $T$ and pressure $P$ fields generated by shock-induced single pore collapse followed by hotspot ignition and growth, is illustrated in the top rows of Figure \ref{fig:micro_prediction} (a) and (b), respectively. Roughly speaking, the time-periods $t \in [0, \, 1.03 \, \mathrm{ns}]$ and $t \in [1.03 \, \mathrm{ns} , \, 2.75 \, \mathrm{ns}]$ represent the above-mentioned two phases of the underlying dynamics, \textit{i.e.}, the ignition and the hotspot growth phases, respectively. The second rows of these figures show the global dynamics of the predicted temperature $\hat{T}$ and pressure field $\hat{P}$, respectively, resulting from the long-term rollout of our autoregressive model, given the initial condition of $T, P, \mu$ at $t=0$. Recall that, in our full model, the predicted temperature field $\hat{T}$ (resp. predicted pressure field $\hat{P}$) is reconstructed by the temperature variational decoder $D_T$ (resp. pressure variational decoder $D_P$) from the evolved temperature latent mean field $\bar{Z}_T$ and the temperature latent log-variance field $Z^\sigma_T$ (resp. pressure latent mean field $\bar{Z}_P$ and the pressure latent log-variance field $Z^\sigma_P$). 

\begin{figure}
     \centering
     \begin{subfigure}{\textwidth}
         \centering
         \includegraphics[width=\linewidth]{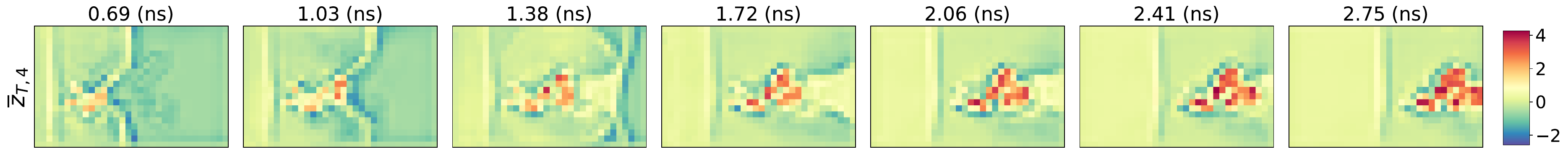}
         \includegraphics[width=\linewidth]{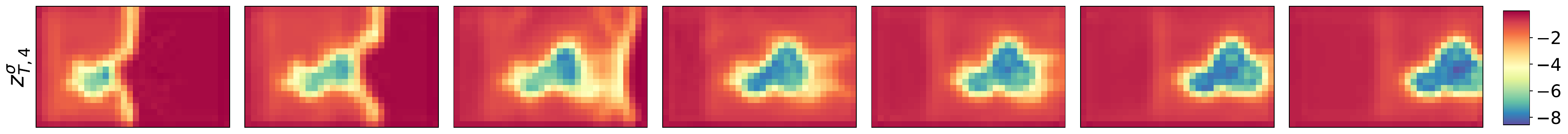}
         \caption{}
     \end{subfigure}
     \begin{subfigure}{\textwidth}
         \centering
         \includegraphics[width=\linewidth]{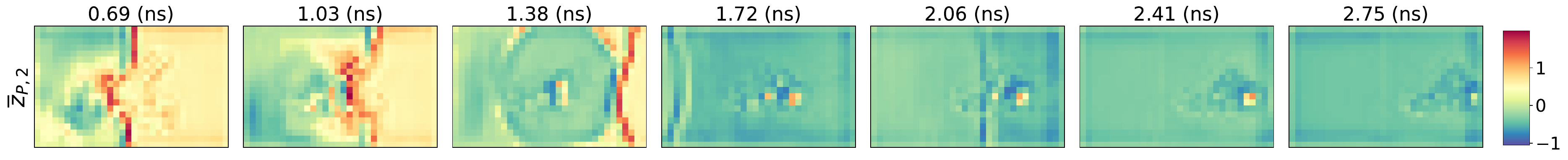}
         \includegraphics[width=\linewidth]{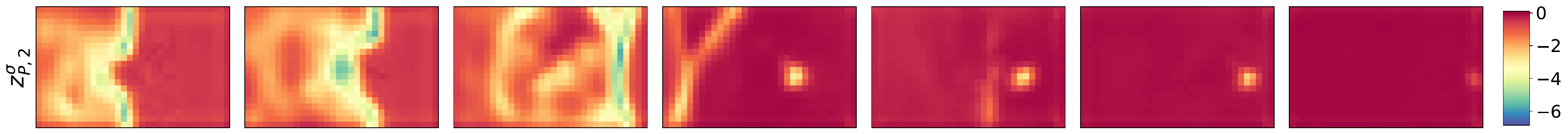}
         \caption{}
     \end{subfigure}
     \caption{Evolution of latent mean and latent log-variance fields. (a) dominant latent field in the prediction of the temperature field $T$. (b) dominant latent field in the prediction of the pressure field $P$.}
     \label{fig:latent_field_evolution}
\end{figure}

In this work, we consider each of the latent mean field $\bar{Z}_U$ and the latent log-variance field $Z^\sigma_U$, $U = T, P , \mu$, as a four-dimensional field over the underlying spatial domain $\mathbb{X}$, \textit{i.e.},
\begin{equation}
    \bar{Z}_U = \left( {\bar{Z}_{U,1}} \,,\, {\bar{Z}_{U,2}} \,,\, {\bar{Z}_{U,3}} \,,\, {\bar{Z}_{U,4}} \right)
\end{equation}
and 
\begin{equation}
    Z^\sigma_U = \left( {Z^\sigma_{U,1}} \,,\, {Z^\sigma_{U,2}} \,,\, {Z^\sigma_{U,3}} \,,\, {Z^\sigma_{U,4}} \right) \, ,
\end{equation}
where each of ${\bar{Z}_{U,j}}$ and ${Z^\sigma_{U,j}}$, $j=1, \cdots, 4$, denotes a scalar field over $\mathbb{X}$. In other words, each latent field $Z_U$ is considered to be a discrete Gaussian random field over $\mathbb{X}$, where, for each $x \in \mathbb{X}$, $Z_U (x)$ is a four-dimensional Gaussian random variable.\footnote{Using the notation introduced in \eqref{eq:encoder_convolution_1} and \eqref{eq:encoder_convolution_2}, we have considered $d=4$. Note that the choice of the dimension of the latent field can vary depending on the considered dataset and the chosen neural network architecture of the VAE.} Figure \ref{fig:latent_field_evolution} (a) and (b) illustrate the evolution of the dominant component among the four components of $(\bar{Z}_T , Z^\sigma_T)$ and $(\bar{Z}_P , Z^\sigma_P)$, respectively. Here, we have considered a pair of scalar fields $(\bar{Z}_{T,j} , Z^\sigma_{T,j})$ (resp. $(\bar{Z}_{P,j} , Z^\sigma_{P,j})$) as a dominant component if it captures the evolution of most of the features which are fundamental for reconstructing the corresponding field $T$ (resp. $P$) in the input space. Note that the dominant components $(\bar{Z}_{T,4} , Z^\sigma_{T,4})$ and $(\bar{Z}_{P,2} , Z^\sigma_{P,2})$)  of the temperature and pressure latent field, respectively, involve patterns induced from \emph{all} the three coupled fields: temperature $T$, pressure $P$ and microstructural morphology $\mu$. This fact highlights the importance of the correlation autoregressive functional $\bar{F}_{\mathrm{corr}}$, which \emph{intertwines} the dominant features from all the three coupled fields that are essential for learning the underlying \emph{global} dynamics. 

\subsection{Learning the correlation between micro-scale building blocks} \label{sec:correlation_scale_bridging}

Following our proposed meta-learning framework, we consider the micro-scale latent fields $(\bar{Z}_T , Z^\sigma_T)$, $(\bar{Z}_P , Z^\sigma_P)$ and $(\bar{Z}_\mu , Z^\sigma_\mu)$, whose learning process is described in Section \ref{sec:micro-dynamics}, as the building blocks of meso-scale dynamics of the coupled fields $(T, P, \mu)$. In this section, we describe how this micro-scale information is leveraged to learn the corresponding meso-scale dynamics by training our ``micro-scale-learned'' model using a relatively small dataset of meso-scale simulations. 

Let $\mathbb{X}_{\mathrm{mes}}$ denote a meso-scale domain of porous energetic material consisting of a field of voids. Consider the parametrized dynamical system 
\begin{equation} \label{dynamical-system-2}
    \frac{d}{dt} \mathbf{U}  = \widetilde{\mathbf{F}}_{\mathrm{mes}} (\mathbf{U}; \psi) \, , \qquad  \psi \in \Psi \, ,    
\end{equation}
corresponding to the evolution of the state vector $\mathbf{U} = (T, P , \mu)$, where the parameters $\psi \in \Psi$ represent the geometrical and morphological characteristics of the underlying domain $\mathbb{X}_{\mathrm{mes}}$ at $t=0$.\footnote{In this work, we restrict ourselves to the case of a fixed energetic material. Hence, the variations between the chemical properties of various energetic material is beyond the scope of this work.} It is known that, depending on the geometrical properties of the underlying field of voids, the patterns of the temperature $T$ and pressure $P$ fields and their evolution, during the hotspot ignition and growth, can be quite different. In other words, the global behavior of the dynamical system \eqref{dynamical-system-2} can change considerably by varying the geometric parameters $\psi$. In the following, we assume that the parametric solution map $\psi \mapsto \mathbf{U}_{\mathrm{mes}}(t; \psi)$ is a well-defined mapping from a low-dimensional manifold in the parameter space $\Psi$ to an appropriate function space over the time-interval $[0, \mathbb{T}]$. In addition, we suppose that the meso-scale microstructures are sampled from a probability distribution $\mathcal{P}(\psi)$ over the parameter space $\Psi$. Using the parametric solution map $\psi \mapsto \mathbf{U}_{\mathrm{mes}}(\cdot \, ; \psi)$, one can push-forward $\mathcal{P}(\psi)$ to the corresponding function space in order to get a stochastic process ${\{ \mathbf{U}_{\mathrm{mes}}(t) \}}_{t \in [0, \mathbb{T}]}$ over the state space of the dynamical system \eqref{dynamical-system-2}. We use our probabilistic model to learn the underlying dynamics corresponding to this stochastic process.

Let ${\{ \mathbf{Z}_{\mathrm{mes}}(t) \}}_{t \in [0, \mathbb{T}]}$ be the stochastic process over the latent space, where $\mathbf{Z}_{\mathrm{mes}} (t)$ denotes the direct sum of the Gaussian random fields encoded by the variational encoders for the temperature $T$, pressure $P$ and microstructural morphology $\mu$ fields, respectively. According to the Bayes' theorem, the marginal probability distributions of the two stochastic processes ${\{ \mathbf{U}_{\mathrm{mes}}(t) \}}_{t \in [0, \mathbb{T}]}$ and ${\{ \mathbf{Z}_{\mathrm{mes}}(t) \}}_{t \in [0, \mathbb{T}]}$satisfy
\begin{equation} \label{Bayes}
    \mathcal{P} \left( {{\mathbf{Z}_{\mathrm{mes}}(t)} \big| {\mathbf{U}_{\mathrm{mes}}(t)}} \right)
    \mathcal{P} \left( {\mathbf{U}_{\mathrm{mes}}(t)} \right)
    = 
    \mathcal{P} \left( {{\mathbf{U}_{\mathrm{mes}}(t)} \big| {\mathbf{Z}_{\mathrm{mes}}(t)}} \right)
    \mathcal{P} \left( {\mathbf{Z}_{\mathrm{mes}}(t)} \right) \, .
\end{equation}
at each $t \in [0, \mathbb{T}]$. The two conditional probability distributions ${\mathcal{P} \left( {{\mathbf{Z}_{\mathrm{mes}}(t)} \big| {\mathbf{U}_{\mathrm{mes}}(t)}} \right)}$ and ${\mathcal{P} \left( {{\mathbf{U}_{\mathrm{mes}}(t)} \big| {\mathbf{Z}_{\mathrm{mes}}(t)}} \right)}$, in \eqref{Bayes}, are approximated by the variational encoder and decoder of our model, respectively. In other words, we use the pre-trained variational autoencoders on the micro-scale data, while their trainable parameters are kept frozen, to encode the input field $\mathbf{U}_{\mathrm{mes}}$ into the corresponding latent mean field ${\bar{\mathbf{Z}}}_{\mathrm{mes}}$ and latent log-variance field ${\mathbf{Z}}^\sigma_{\mathrm{mes}}$, and decode back into the input space. Hence, in order to learn the evolution of the state vector $\mathbf{U}_{\mathrm{mes}} = (T, P, \mu)$ corresponding to a meso-scale domain, it suffices to learn the dynamics of the latent Gaussian random field ${\mathbf{Z}_{\mathrm{mes}}(t)}$. 

We consider ${\{ \mathbf{Z}_{\mathrm{mes}}(t) \}}_{t \in [0, \mathbb{T}]}$ as a Markov process whose transition operator $\widehat{\mathbf{F}}_{\mathrm{mes}}$ describes the evolution of the corresponding marginal probability distribution, \textit{i.e.}, 
\begin{equation}
    \mathcal{P} \left( {{\mathbf{Z}}_{|t_{j+1}}} \right) =
    \widehat{\mathbf{F}}_{\mathrm{mes}} \, \mathcal{P} \left( {\mathbf{Z}_{|t_{j}}} \right) \, .
\end{equation}
In addition, since $\mathbf{Z}_{\mathrm{mes}}(t)$ is considered to be a discrete Gaussian random field with trivial auto-correlation function with respect to the underlying spatial coordinate, the Markov transition operator $\widehat{\mathbf{F}}_{\mathrm{mes}}$ can be reformulated in terms of an autoregressive functional $\mathbf{F}_{\mathrm{mes}}$ acting on the corresponding mean and log-variance fields in the following form  
\begin{equation} \label{autoregression-meso}
    {\left( {\bar{\mathbf{Z}} , {\mathbf{Z}}^{\sigma}} \right)}_{|t_{j+1}} =
    \mathbf{F}_{\mathrm{mes}} \left(
    {\left( {\bar{\mathbf{Z}} , {\mathbf{Z}}^{\sigma}} \right)}_{|t_{j}} 
    \right) \, , 
\end{equation}
as mentioned in \eqref{autoregression-1}. The detailed way in which $\mathbf{F}_{\mathrm{mes}}$ acts on each component of $\bar{\mathbf{Z}}$ and ${\mathbf{Z}}^{\sigma}$ is described in Equations \eqref{deterministic-dynamics-1} - \eqref{stochastic-dynamics-3}. We formulate the problem of learning the autoregressive functional $\mathbf{F}_{\mathrm{mes}}$ in a meta-learning framework in the following.

As discussed before, generating the simulation data for the evolution of the state vector $\mathbf{U} = (T, P , \mu)$ in the case of micro-scale single pore collapse is much cheaper than the corresponding meso-scale simulation. Hence, in the first step, we learn the micro-scale autoregressive functional $\mathbf{F}_{\mathrm{mic}}$ over a relatively large dataset of single-void collapse simulations. The functional $\mathbf{F}_{\mathrm{mic}}$ captures the coupled evolution of the temperature $T$, pressure $P$ and microstructural morphology $\mu$ fields in a micro-scale domain. We consider $\mathbf{F}_{\mathrm{mic}}$ as a ``first-order approximation'' of the meso-scale autoregressive functional $\mathbf{F}_{\mathrm{mes}}$ in the following sense. Motivated by tokenization in Natural Language Processing (NLP), suppose the meso-scale domain $\mathbb{X}_{\mathrm{mes}}$ can be decomposed into a disjoint union of micro-scale domains $\mathbb{X}^i_{\mathrm{mic}}$, $i \in I$, as
\begin{equation}
    \mathbb{X}_{\mathrm{mes}} = \bigsqcup_{i \in I} \mathbb{X}^i_{\mathrm{mic}}\, ,
\end{equation}
in such a way that the morphological characteristics of each micro-scale domain $\mathbb{X}^i_{\mathrm{mic}}$ is represented by a parameter $\psi^i \in \Psi$ in the parameter space $\Psi$. For each micro-scale domain $\mathbb{X}^i_{\mathrm{mic}}$, $i \in I$, let $\mathbf{U}^i_{\mathrm{mic}}$ be the evolution of the state vector $\mathbf{U} = (T, P, \mu)$ over $\mathbb{X}^i_{\mathrm{mic}}$ which is predicted by the micro-scale trained model $\mathbf{F}_{\mathrm{mic}}$ given the data ${\left( \mathbb{X}^i_{\mathrm{mic}}, \psi^i \right)}$. After applying the appropriate time-delay necessary for the shock wave propagation over a meso-scale domain, the direct sum of the predicted micro-scale evolutions $\mathbf{U}^i_{\mathrm{mic}}$ might be considered as a rough approximation of $\mathbf{U}_{\mathrm{mes}}$ for a very short period of time-span. However, it is known that the interaction between different voids in a meso-scale domain play a crucial role in the long time-span of meso-scale dynamics $\mathbf{U}_{\mathrm{mes}}$. Hence, our micro-scale autoregressive model $\mathbf{F}_{\mathrm{mic}}$ needs to learn the correlation between the micro-scale building blocks of the underlying meso-scale dynamics $\mathbf{U}_{\mathrm{mes}}$ in order to capture the fundamental effects of void-void interaction at meso-scale. In other words, one can consider 
\begin{equation} \label{eq:higher_order_correction}
    \mathbf{F}_{\mathrm{mes}} = \mathbf{F}_{\mathrm{mic}} + \text{``higher-order correction terms''} \,
\end{equation}
where the correction terms capture the mutual interaction between the dynamics of neighboring micro-scale latent fields.

\begin{figure}
    \centering
    \includegraphics[width=0.3\linewidth]{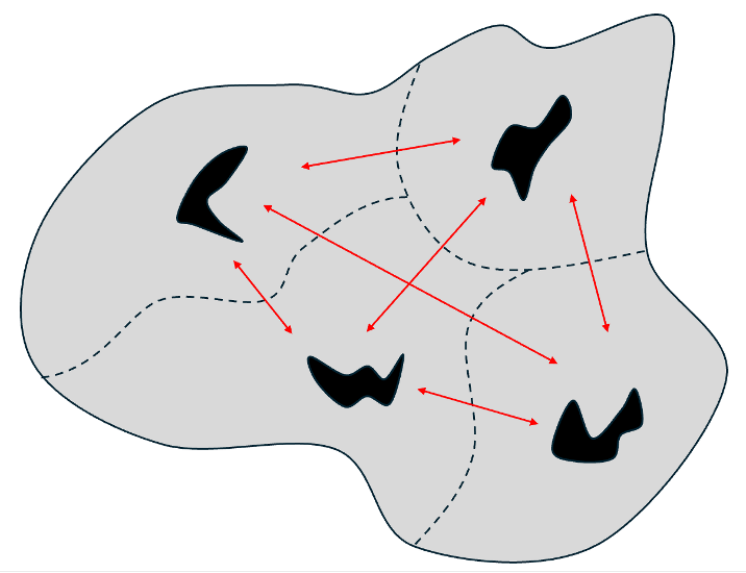}
    \caption{Division of a meso-scale domain $\mathbb{X}_{\mathrm{mes}}$ into micro-scale subdomains $\mathbb{X}^i_{\mathrm{mic}}$: void-void interaction at the meso-scale}
\end{figure}

The problem of solving a parametrized family of PDEs can be seen through the lens of meta-learning as discussed in \citep{penwarden2023metalearning}. In this setup, one considers learning the PDE solution corresponding to each region of the parameter space over which the mathematical characteristics of the PDE remain the same, \textit{e.g.}, the Navier-Stokes equation with specific Reynolds number range, as a \emph{task}. Hence, learning the PDE solution corresponding to the non-overlapping subsets of the full parameter space can be considered as a family of closely related tasks which one aims to extract common meta information from them. Similarly, we consider the problem of learning the micro-scale dynamics $\mathbf{F}_{\mathrm{mic}}$ of the state vector $\mathbf{U}=(T,P, \mu)$ corresponding to the collapse of single-voids with various morphological characteristics $\psi \in \Psi$ as a family of correlated tasks. The common knowledge extracted from this family of tasks gives a short-term approximation of the dynamics over each micro-scale domain $\mathbb{X}^i_{\mathrm{mic}}$. We use this fundamental information by transferring our micro-scale trained model to a meso-scale domain, and using its learned parameters as the initial-value for the trainable parameters of the meso-scale dynamics model $\mathbf{F}_{\mathrm{mes}}$. 

As mentioned before, we consider a fully convolutional, \textit{i.e.}, CNN-based, architecture for the autoregressive mapping in our model which enables us to transfer the model from micro- to meso-scale. Let $\bar{\mathbf{L}}_\theta$ and ${\mathbf{L}}^\sigma_{\theta'}$ be the convolution kernels corresponding to the deterministic $\bar{\mathbf{F}}$ and stochastic ${\mathbf{F}}^\sigma$ parts of the latent evolution model given by \eqref{deterministic-dynamics-1} and \eqref{stochastic-dynamics-1}, respectively. That is
\begin{align}
    {\bar{\mathbf{Z}}}_{|t_{j+1}} (x) 
    &= \bar{\mathbf{F}} \left( {\bar{\mathbf{Z}}}_{|t_{j}} \right) (x) \\ \nonumber
    &= \left( {\Bar{\mathbf{L}}_{\theta} \ast {\bar{\mathbf{Z}}}_{|t_{j}}} \right) (x) \\ \nonumber
    &= \int_{\mathbb{X}_{\mathrm{mes}}} {\Bar{\mathbf{L}}_{\theta} (x-y) \cdot {\bar{\mathbf{Z}}}_{|t_{j}}(y)} \, \mathrm{d} y 
    \, , \qquad x \in \mathbb{X}_{\mathrm{mes}} \, ,
\end{align}
and 
\begin{align}
    {{\mathbf{Z}}^\sigma_{|t_{j+1}}} (x) 
    &= \mathbf{F}^\sigma \left(
    {\left( {\bar{\mathbf{Z}} \oplus {\mathbf{Z}}^{\sigma}} \right)}_{|t_{j}} \right) (x) \\ \nonumber
    &= \left( {{\mathbf{L}}^\sigma_{\theta'} \ast {\left( {\bar{\mathbf{Z}} \oplus {\mathbf{Z}}^{\sigma}} \right)}_{|t_{j}} } \right) (x) \\ \nonumber
    &= \int_{\mathbb{X}_{\mathrm{mes}}} {{\mathbf{L}}^\sigma_{\theta'} (x-y) \cdot {\left( {\bar{\mathbf{Z}} \oplus {\mathbf{Z}}^{\sigma}} \right)}_{|t_{j}} (y) } \, \mathrm{d} y 
    \, , \qquad x \in \mathbb{X}_{\mathrm{mes}} \, .
\end{align}
Considering the notation introduced in \eqref{eq:encoder_convolution_1} and \eqref{eq:encoder_convolution_2}, the convolution kernels $\bar{\mathbf{L}}_\theta$ and ${\mathbf{L}}^\sigma_{\theta'}$ can be considered as matrix-valued functions defined over the meso-scale domain $\mathbb{X}_{\mathrm{mes}}$ with values in $\mathcal{M}_{d \times d}$ and $\mathcal{M}_{d \times 2d}$, respectively. We start the learning process of the trainable parameters $(\theta, \theta')$ of the convolution kernels $\Bar{\mathbf{L}}_{\theta}$ and ${\mathbf{L}}^\sigma_{\theta'}$ by training our model on a relatively large dataset of inexpensive micro-scale simulations. After the model learns a latent representation of the micro-scale dynamics, we transfer it and train it on a small dataset of meso-scale simulations in order to learn the correlations between neighboring micro-scale building blocks of the meso-scale dynamics. In other words, the learned parameters ${(\theta, \theta')}_{\mathrm{mic}}$ at the micro-scale are considered as the initialization of the meso-scale model's parameters ${(\theta, \theta')}_{\mathrm{mes}}$, and they will be updated as the model learns the higher-order corrections mentioned in \eqref{eq:higher_order_correction}. Roughly speaking, the transition from the initial parameters ${(\theta, \theta')}_{\mathrm{mic}}$ to ${(\theta, \theta')}_{\mathrm{mes}}$ in our model can be considered as a way of ``stitching'' the micro-scale dynamics over neighboring micro-scale domains $\mathbb{X}_{\mathrm{mic}}^i$ using the convolutional formulation of the latent evolution autoregressive functional $\mathbf{F}_{\mathrm{mes}}$. Considering our CNN-based architecture, the above-mentioned higher-order corrections take into account the effects of short-range void-void interaction in each \emph{local} neighborhood of $\mathbb{X}_{\mathrm{mes}}$ which is covered by the \emph{receptive field} \citep{luo2016understanding} of the convolution kernels $\Bar{\mathbf{L}}_{\theta}$ and ${\mathbf{L}}^\sigma_{\theta'}$.\footnote{The notion of receptive field of convolutional neural networks is a important characteristic of CNNs which represent the pixel domain in the input image which affects the value of each output pixel, and the area outside of the receptive field has no contribution to the corresponding output pixel. Mathematically, it is equivalent to the notion of \emph{support} of the function $\Bar{\mathbf{L}}_{\theta} (x,y)$ (or ${\mathbf{L}}^\sigma_{\theta'} (x,y)$) for each fixed $x \in \mathbb{X}_{\mathrm{mes}}$, which is defined as the set of $y \in \mathbb{X}_{\mathrm{mes}}$ for which $\Bar{\mathbf{L}}_{\theta}(x,y) \neq 0$.} 

\begin{figure}
     \centering
     \begin{subfigure}{\textwidth}
         \centering
         \includegraphics[width=\linewidth]{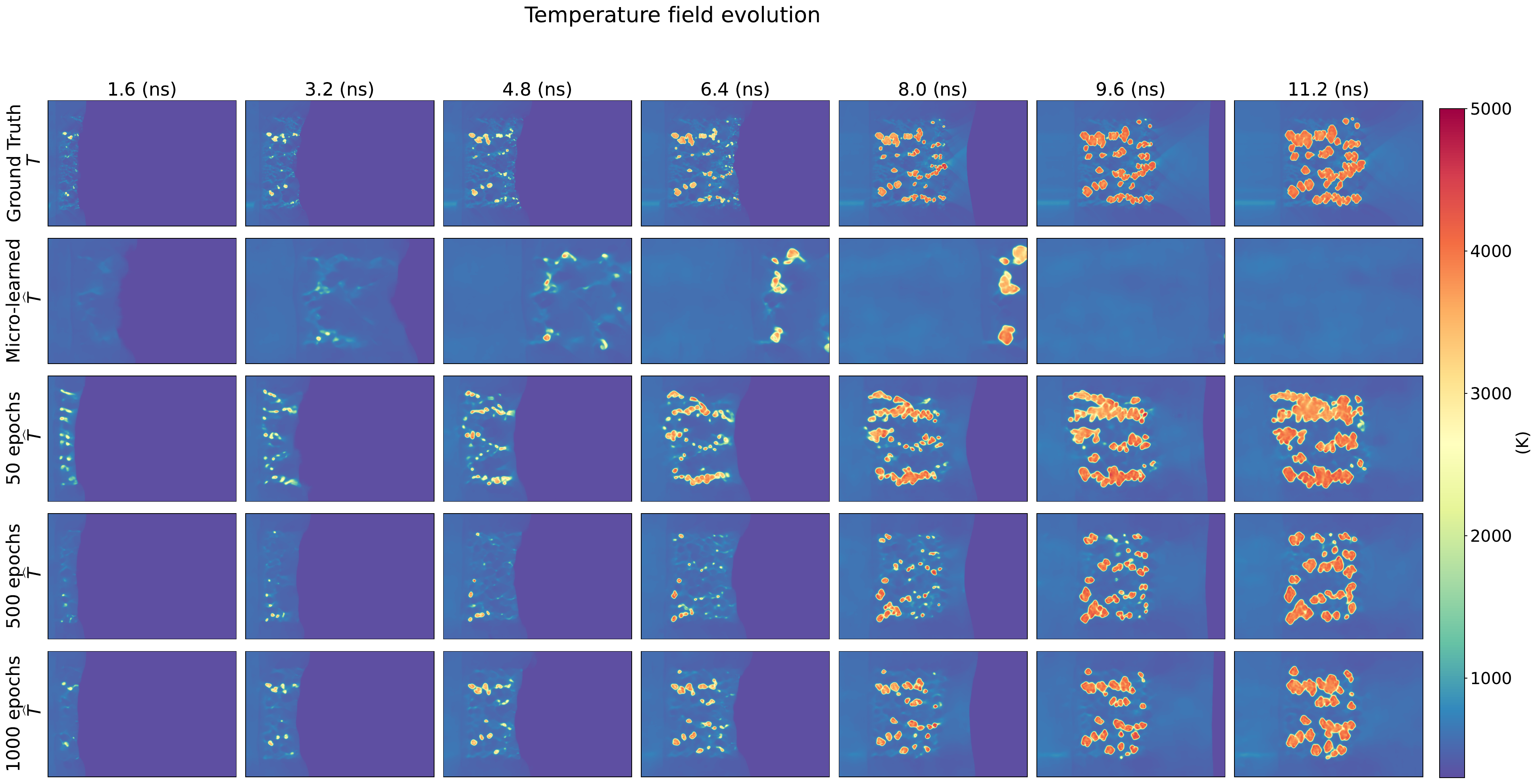}
         \caption{}
     \end{subfigure}
     \begin{subfigure}{\textwidth}
         \centering
         \includegraphics[width=\linewidth]{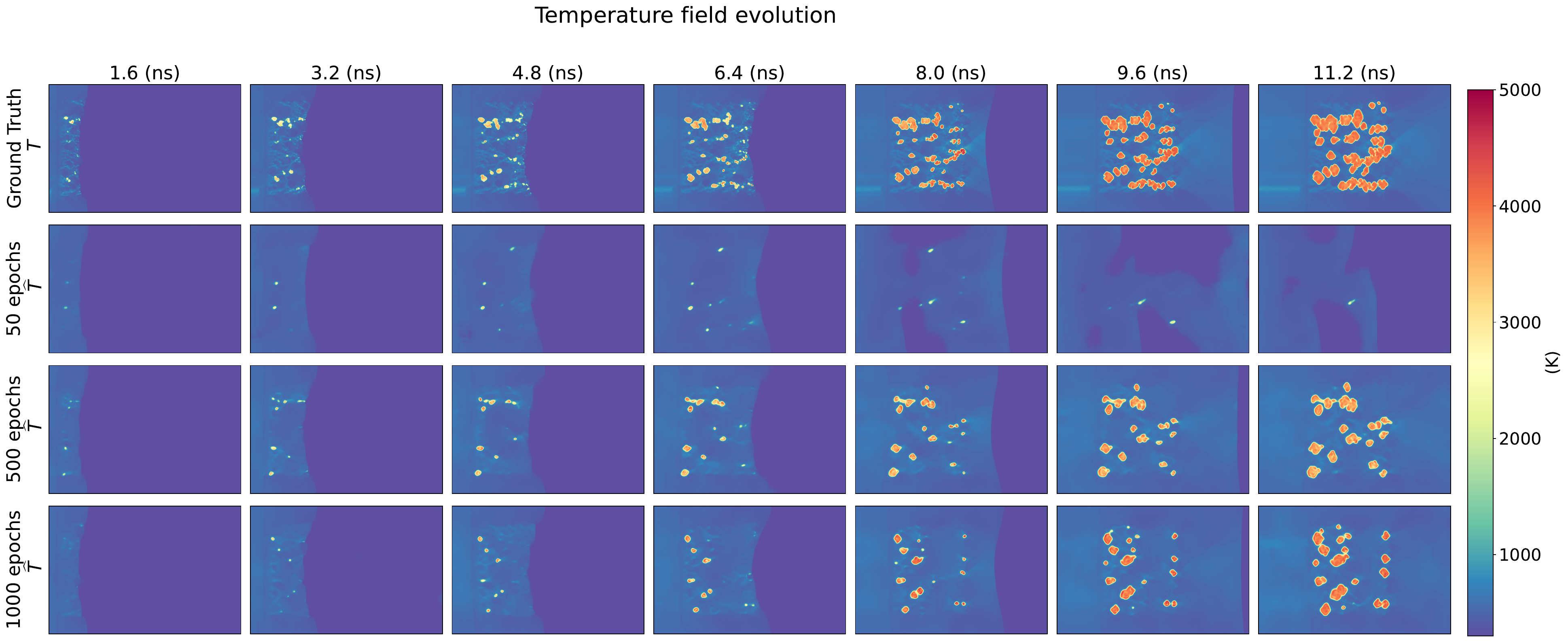}
         \caption{}     
     \end{subfigure}
     \caption{Predicted meso-scale temperature evolution corresponding to progressive training of meso-scale autoregressive functional $\mathbf{F}_{\mathrm{mes}}$  (a) initialized from micro-scale-learned model $\mathbf{F}_{\mathrm{mic}}$ (b) initialized with random weights from scratch.}
     \label{fig:meso_before_after_correlation}
\end{figure}

Figure \ref{fig:meso_before_after_correlation} (a) illustrates the progressive improvement in the predicted meso-scale temperature evolution as the latent autoregressive model undergoes a transition from $\mathbf{F}_{\mathrm{mic}}$ to $\mathbf{F}_{\mathrm{mes}}$. The first row shows the ground-truth, and the second row corresponds to the predicted evolution by the micro-scale-trained autoregressive functional $\mathbf{F}_{\mathrm{mic}}$ before learning the mutual interaction between the dynamics of neighboring micro-scale building blocks. The third up to the fifth row illustrate the predicted evolution by $\mathbf{F}_{\mathrm{mes}}$ after training for 50, 500, and 1,000 epochs, respectively, over the meso-scale dataset. It can be seen that the deviation between the ground-truth and the $\mathbf{F}_{\mathrm{mic}}$-predicted temperature evolution becomes more considerable for longer time spans. This observation is consistent with our hypothesis in which we consider the direct sum of the latent representation of the micro-scale dynamics to approximate the meso-scale dynamics \emph{only} over a very short time-period. After only 50 epochs of training over meso-scale dataset, the updated model is able to capture the main features of the global dynamics at the meso-scale in terms of the overall pattern of hotspots in addition to the propagation of the shock-front. However, there is a considerable false-positive predicted ignition after 50 epochs of training, which is mainly due to the fact that all the micro-scale single-pore collapse cases seen by the model has led to hotspot ignition, which is not the case at the meso-scale. As the training over the meso-scale dataset continues, \textit{i.e.}, after 500 and 1000 epochs, the prediction performance of the meso-scale model $\mathbf{F}_{\mathrm{mes}}$ improves by learning the higher-order corrections which capture the correlation between neighboring micro-scale building blocks. 

In order to highlight the crucial role played by the first step in our two-step meta-learning framework, \textit{i.e.}, learning $\mathbf{F}_{\mathrm{mic}}$ as a ``first order approximation'' of $\mathbf{F}_{\mathrm{mes}}$, we also consider a setup in which the meso-scale autoregressive functional $\mathbf{F}_{\mathrm{mes}}$ is trained from \emph{scratch} over the same meso-scale dataset considered above. In other words, instead of considering the micro-scale-learned parameters ${(\theta, \theta')}_{\mathrm{mic}}$ as the initial value for ${(\theta, \theta')}_{\mathrm{mes}}$, the trainable parameters ${(\theta, \theta')}_{\mathrm{mes}}$ of the meso-scale autoregressive functional $\mathbf{F}_{\mathrm{mes}}$ gets initialized randomly. The second up to the fourth row of Figure \ref{fig:meso_before_after_correlation} (b) illustrate the predicted temperature evolution by $\mathbf{F}_{\mathrm{mes}}$ after training for 50, 500, and 1,000 epochs, respectively, from scratch. One can see that, even after 1000 epochs of training $\mathbf{F}_{\mathrm{mes}}$ from scratch, the model prediction misses a considerable portion of hotspot ignition and growth compared to the ground-truth.

\subsection{Comparison with a Physics-aware Recurrent Convolutional neural network (PARC) in scarce meso-scale data regime} \label{sec:compare_parc_qualitative}

In this section, we compare the prediction performance of our model with a Physics-aware Recurrent Convolutional neural network (PARC)  \citep{nguyen2023parc,nguyen2025parcv2} in a scarce meso-scale data regime. The choice of PARC, among other Physics-Informed Machine Learning (PIML) \citep{karniadakis2021physics} models, is motivated by the underlying \emph{inductive bias} considered in its architecture design which is particularly suited for learning fast and transient spatiotemporal dynamics with sharp gradients (see \citep{nguyen2025parcv2} for comparison between prediction performance of PARCv2 and other PIML models in the case of micro-scale single-pore collapse in EM).\footnote{Note that because of lack of a closed system of PDEs governing the evolution of the physical fields $(T, P)$ in shock-induced energy localization in EMs, the PIML models which are based on the \emph{learning bias} approach are not applicable to this problem.} PARC has a differentiator-integrator architecture, which learns the evolution of physical fields represented by a dynamical system of the form
\begin{equation} 
    \frac{d}{dt} \mathbf{U}  = \widetilde{\mathbf{F}} (\mathbf{U}; \psi) \, , \qquad t \in [0, \mathbb{T}] \, , \quad \psi \in \Psi \, .    
\end{equation}
The differentiator module aims to learn the infinitesimal evolution of the corresponding physical fields encapsulated into $\widetilde{\mathbf{F}}$ while the integrator module computes $\Delta \mathbf{U}$ between two consecutive time-steps $[t_i , t_i + \Delta t]$. The integrator module is considered to be a classical numerical integrator, \textit{e.g.}, Runge-Kutta methods, together with a neural network learning the higher-order corrections. In addition, to better capture the advection-dominant phenomenon, the PARCv2 differentiator architecture involves an inductive bias which is motivated by the generalized advection-diffusion-reaction equation.

\begin{figure}
    \centering
    \includegraphics[width=0.7\linewidth]{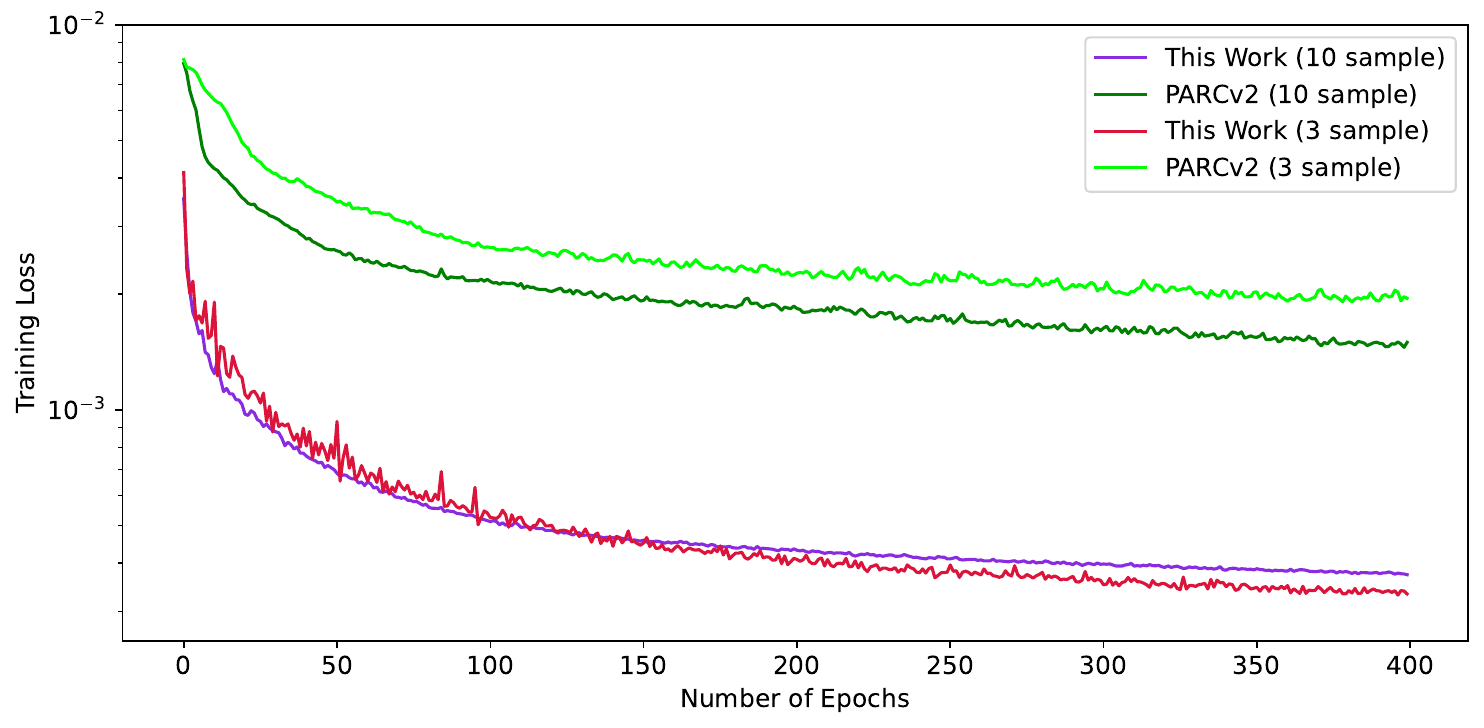}
    \caption{Comparison between the training loss of our model vs. PARCv2 trained on the same meso-scale dataset.}
    \label{fig:training_loss_comparison}
\end{figure}

In order to investigate the performance of both models in the scarce meso-scale data regime, we use a dataset containing 13 instances of shock-initiated reaction simulations in meso-scale microstructures. We consider four training scenarios for each model, using 10, 5, 3 and 1 training samples, respectively. Afterwards, the performance of each model is evaluated on 3 test samples kept separately during the training. Following our two-step meta-learning approach, our model has been trained, initially, on 100 samples of inexpensive micro-scale single-pore collapse simulations. However, PARCv2 is trained from scratch only on the meso-scale data. Figure \ref{fig:training_loss_comparison} compares the convergence rate of two models in terms of their training loss in the case of 10 and 3 meso-scale training samples, respectively. It can be seen that pretraining our model on the micro-scale data allows it to start from a lower initial loss value and converge relatively quickly compared to PARCv2, which is trained from scratch using only the meso-scale simulations. Note that the PARCv2 training loss is closely related to the $\mathcal{L}_{\mathrm{multi-step}}$ loss term, given by \eqref{loss-multi-step}, considered for training our model in the sense that they both compare the predicted fields pixel-wise in the input space. However, we consider two extra loss terms $\mathcal{L}_{\mathrm{consistency-mean}}$ and $\mathcal{L}_{\mathrm{consistency-var}}$, given by \eqref{loss-consistency-mean} and \eqref{loss-consistency-var}, respectively, in training our autoregressive model which compare the predicted dynamics over latent space with the encoded dynamics using pretrained variational encoders.

\begin{figure}
     \centering
     \begin{subfigure}{\textwidth}
         \centering
         \includegraphics[width=\linewidth]{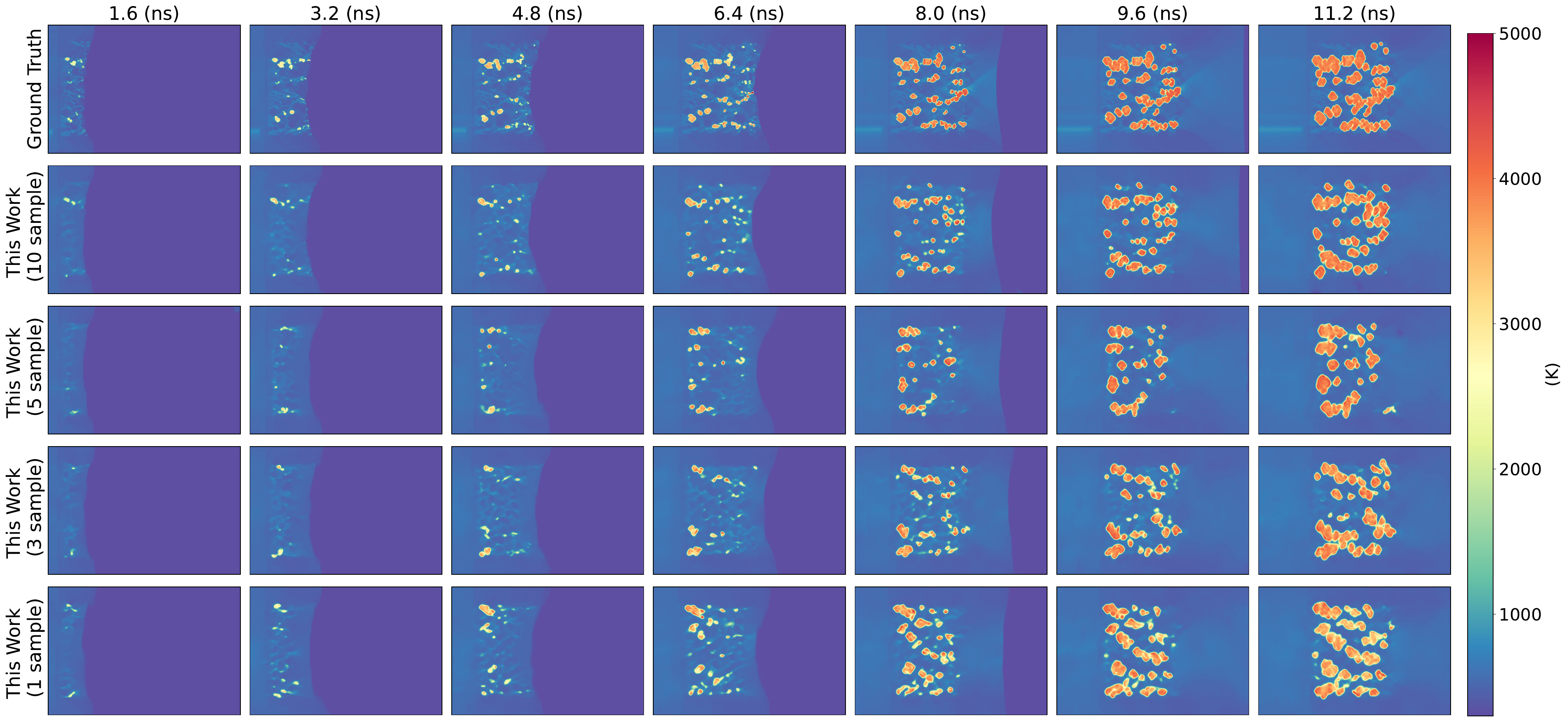}
         \caption{}         
     \end{subfigure}
     \begin{subfigure}{\textwidth}
         \centering
         \includegraphics[width=\linewidth]{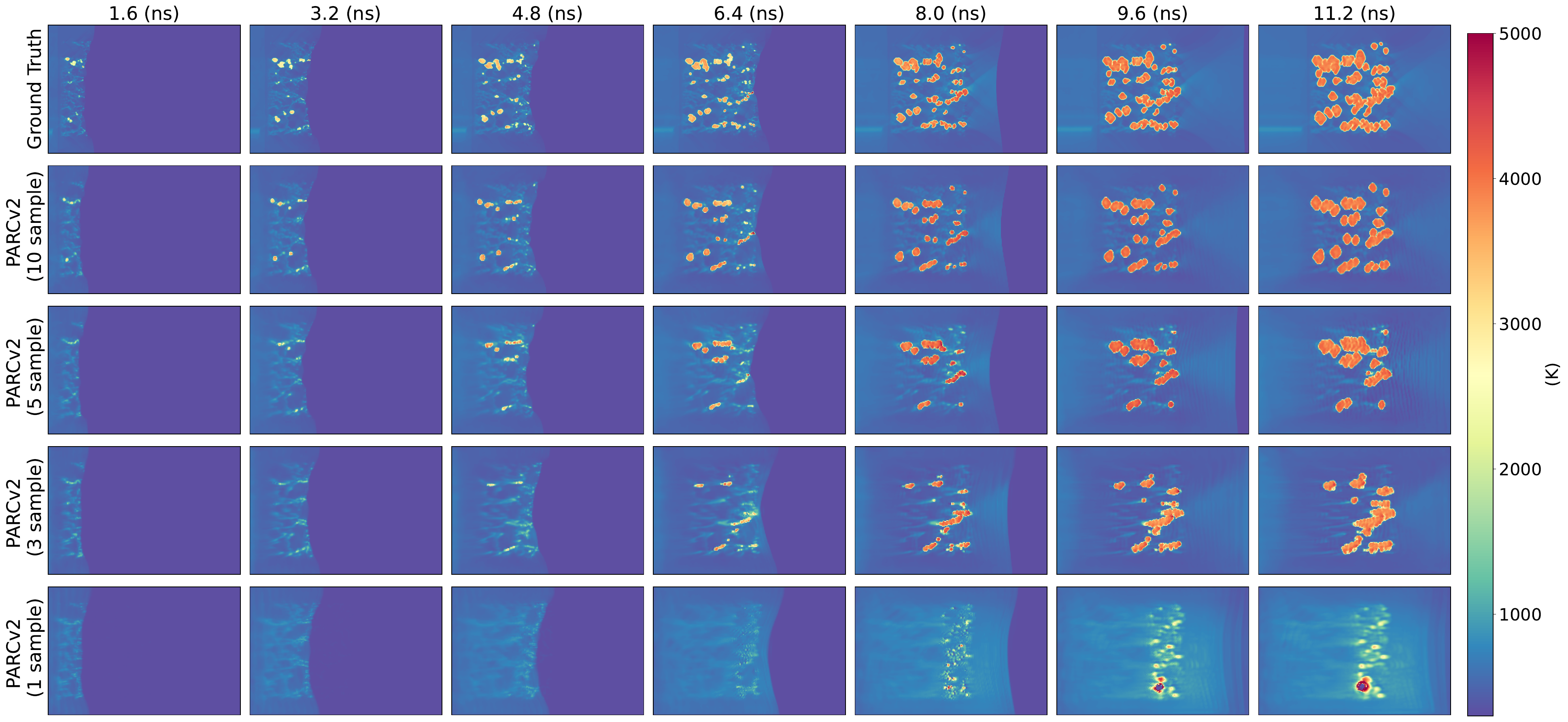}
         \caption{}
     \end{subfigure}
     \caption{Test sample \#1: Comparison between the predicted temperature field evolution by (a) our model and (b) PARCv2 in four training scenarios.}    
     \label{fig:LP_Pv2_comparison_T_prediction}
\end{figure}

\begin{figure}
     \centering
     \begin{subfigure}{\textwidth}
         \centering
         \includegraphics[width=\linewidth]{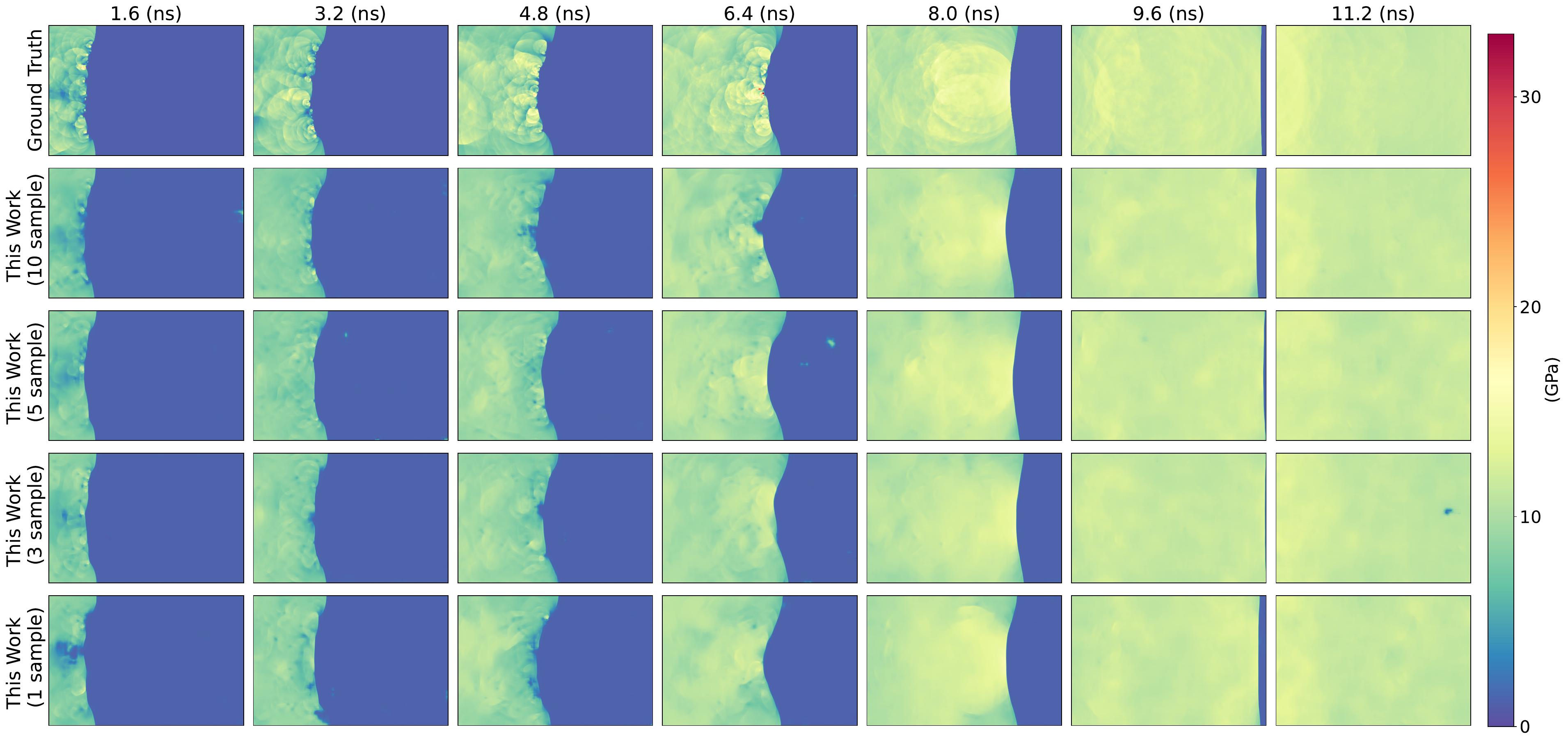}
         \caption{}         
     \end{subfigure}
     \begin{subfigure}{\textwidth}
         \centering
         \includegraphics[width=\linewidth]{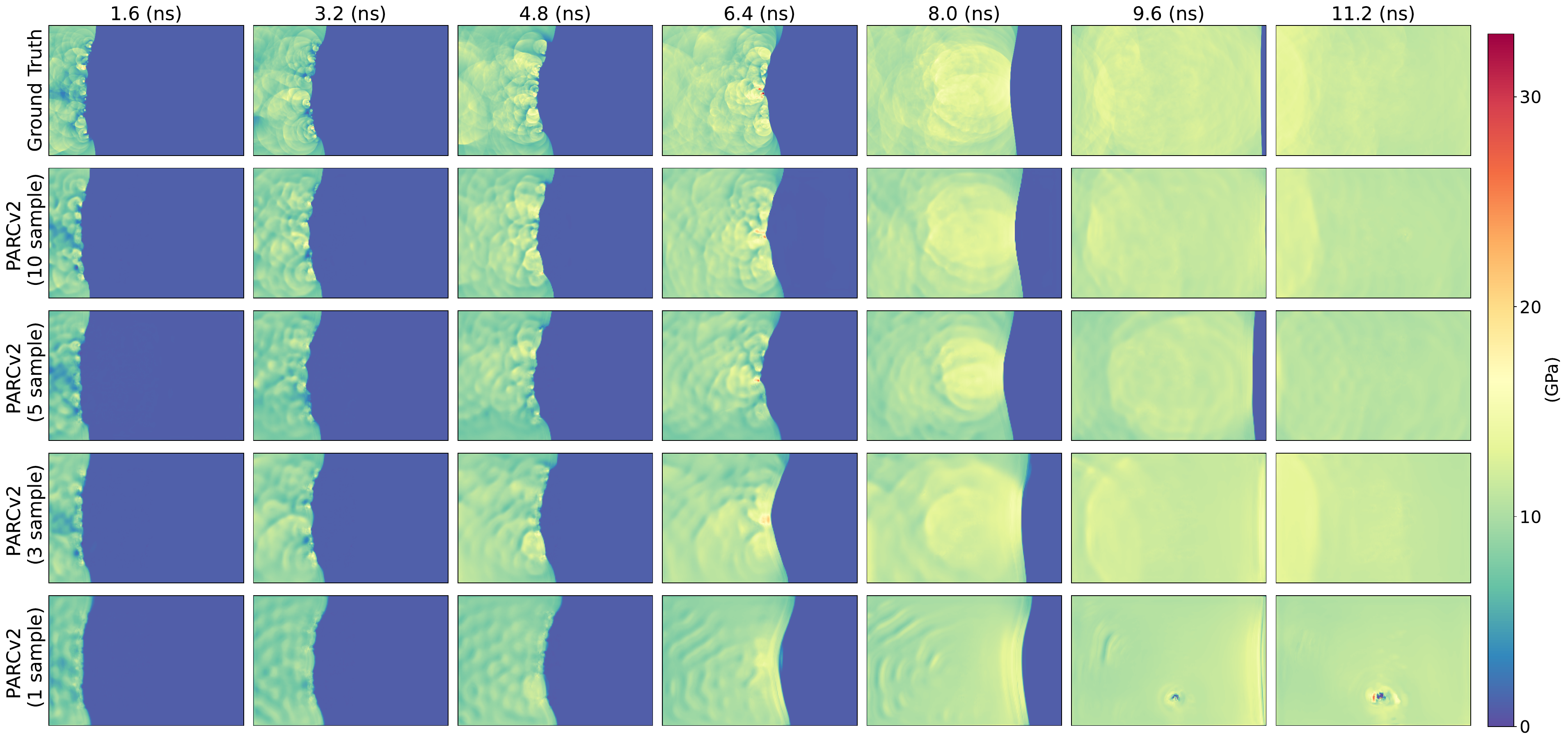}
         \caption{}
     \end{subfigure}
     \caption{Test sample \#1: Comparison between the predicted pressure field evolution by (a) our model and (b) PARCv2 in four training scenarios.}    
     \label{fig:LP_Pv2_comparison_P_prediction}
\end{figure}

Figure \ref{fig:LP_Pv2_comparison_T_prediction} (a) and (b) display the predicted evolution of the temperature field $T$ by our model and PARCv2, respectively, for a test sample in comparison with the ground-truth numerical simulation. The columns in the figures show the predicted temperature distribution at different instants of time (noted above the panels); the rows correspond to the above-mentioned four scenarios considered for training of each model. The first row shows the ground truth, and the second up to the fifth row illustrate the prediction of each model trained on 10, 5, 3, and 1 meso-scale samples, respectively. In terms of capturing the formation and growth of the hotspots, our model's prediction outperforms PARCv2, especially in the last three scenarios with few training samples. Our model's predicted temperature values for the hotspots approaches closer to the ground-truth as the number of training samples increases, but it mostly stays below the ground-truth temperatures. On the other hand, PARCv2 tends to predict fewer hotspots with overpredicted temperature values. The boundary of our model's predicted hotspots is blurrier compared with PARCv2, \textit{i.e.}, it has less sharp gradients relative to the ambient field. Also, similar observations hold in comparing the predicted evolution of the pressure field $P$ by our model and PARCv2 as illustrated in Figure \ref{fig:LP_Pv2_comparison_P_prediction} (a) and (b), respectively. PARCv2 can better capture the delicate patterns of the shock-waves in the pressure field compared to the present model which can be due to the fact that PARCv2 learns the dynamics in the high-fidelity input space instead of in the VAE-compressed latent space. The comparison between the predicted evolution of the temperature $T$ and pressure $P$ fields by our model and PARCv2 in the case of other test samples are presented in Appendix \ref{sec:appendix_pred_test} which confirm similar observations to the above-mentioned test sample. We extend the comparison between the two models' performance in terms of quantitative metrics in the following Subsection.

\subsection{Quantitative evaluation of prediction performances}

In order to further assess the prediction performance of the two models quantitatively, we consider several EM sensitivity metrics which play a crucial role in the design of energetic materials \citep{sen2018multi}. These quantities of interest (QoI) are defined based on the distribution of the temperature field $T(x,t)$ over the material domain together with its temporal evolution. Denote by $\mathbb{D}_\hs \subset \mathbb{X}$ the hotspot domain defined as the subregion of the material domain $\mathbb{X}$ where the temperature of the material exceeds the bulk temperature value $T_{\mathrm{bulk}}$ after the passage of a planar shock wave, \textit{i.e.},
\begin{equation}
    \mathbb{D}_\hs = \left\{ x \in \mathbb{X} \, \big| \, T(x) > T_{\mathrm{bulk}} \right\} \, .
\end{equation}
Following \citep{seshadri2022meso,nguyen2023parc}, we consider $T_{\mathrm{bulk}} = 875 \, \mathrm{K}$ in this work. One of the important QoI for determining the sensitivity of an energetic material is the \emph{total hotspot area} $A_\hs$, which is defined as the area of the hotspot domain $\mathbb{D}_\hs$, \textit{i.e.},
\begin{equation}
    A_\hs = \int_{\mathbb{D}_\hs}  \mathrm{d}x \, .
\end{equation}
It captures the total contribution of hotspots to shock-induced energy localization in a representative volume element of EMs. The second QoI is the \emph{average hotspot temperature} $\bar{T}_\hs$, defined by
\begin{equation}
    \bar{T}_\hs = \frac{1}{A_\hs} \int_{\mathbb{D}_\hs} T(x) \, \mathrm{d}x \, ,
\end{equation}
which encapsulates the average intensity of localized energy and the likelihood of formation of ``critical'' hotspots \citep{tarver1996critical}. In addition, the rate of change of the above-mentioned metrics, denoted by $\dot{A}_\hs$ and ${\dot{\bar{T}}}_\hs$, respectively, are two other important QoI considered in this work.

\begin{figure}
     \centering
     \begin{subfigure}{\textwidth}
         \centering
         \includegraphics[width=0.95\linewidth]{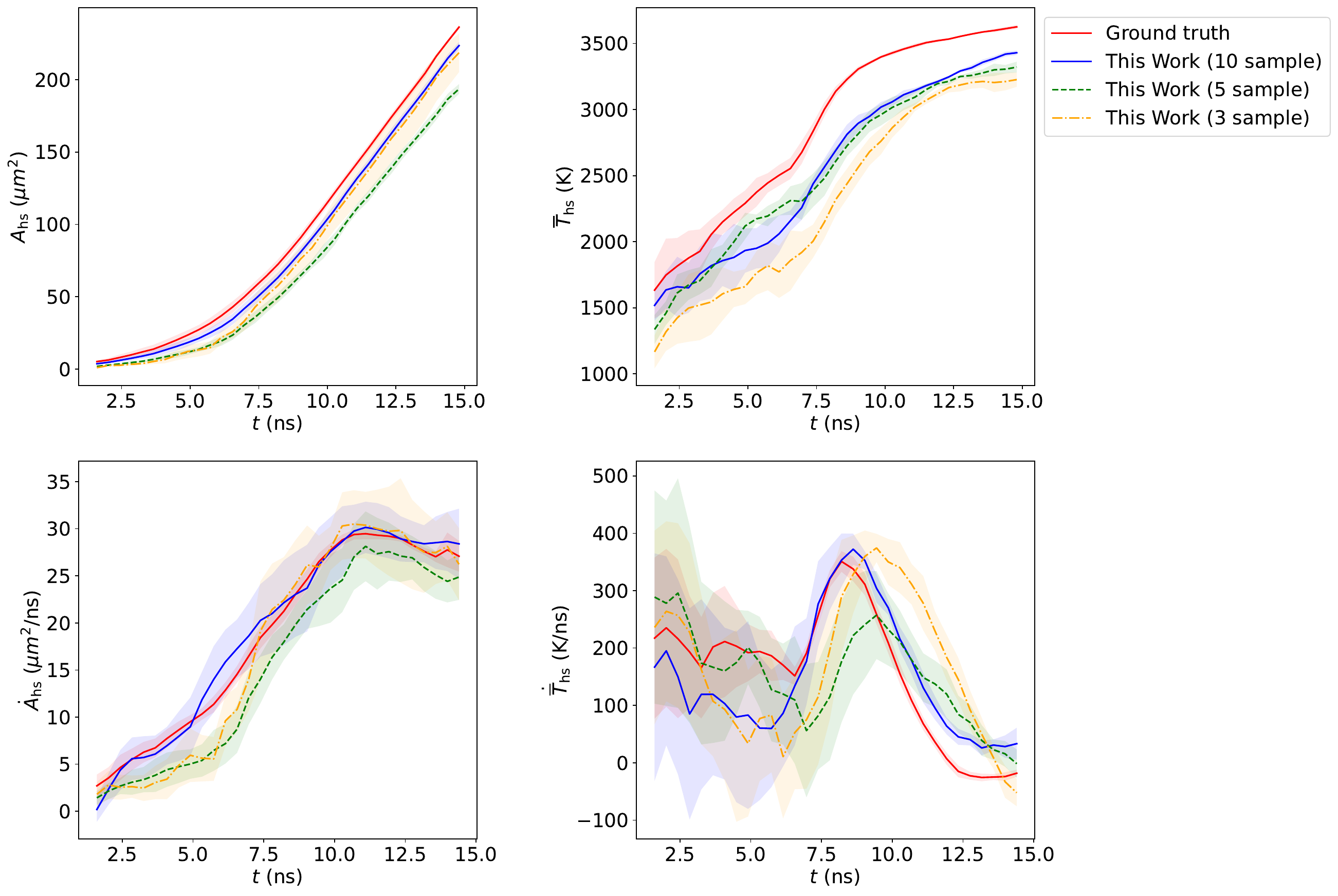}
         \caption{}         
     \end{subfigure}
     \begin{subfigure}{\textwidth}
         \centering
         \includegraphics[width=0.95\linewidth]{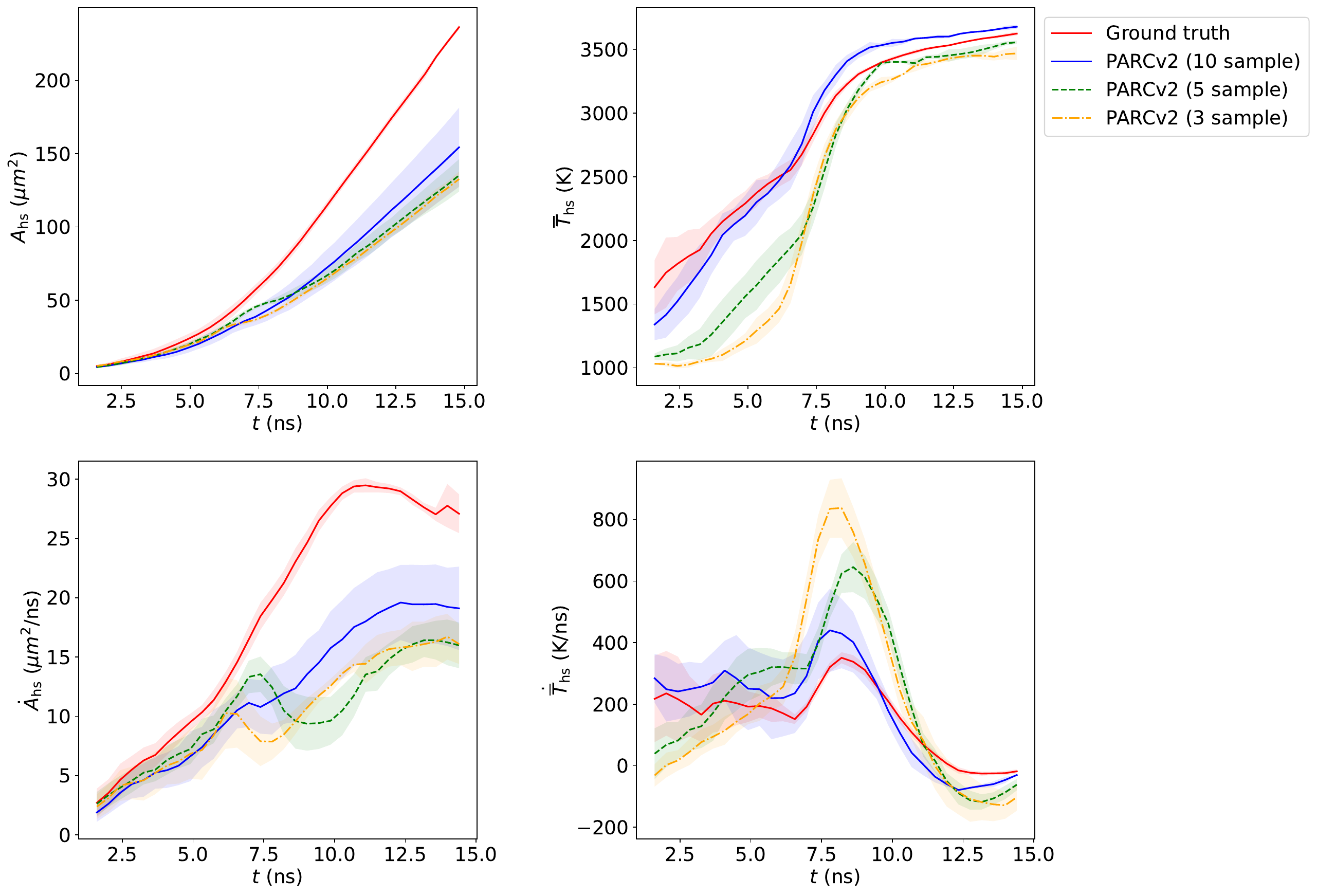}
         \caption{}
     \end{subfigure}
     \caption{EM sensitivity QoI calculated from (a) our model and (b) PARCv2 prediction in comparison with ground-truth.}  
     \label{fig:LP_Pv2_hotspot_metrics}
\end{figure}

\begin{figure}
    \centering
    \includegraphics[width=0.9\linewidth]{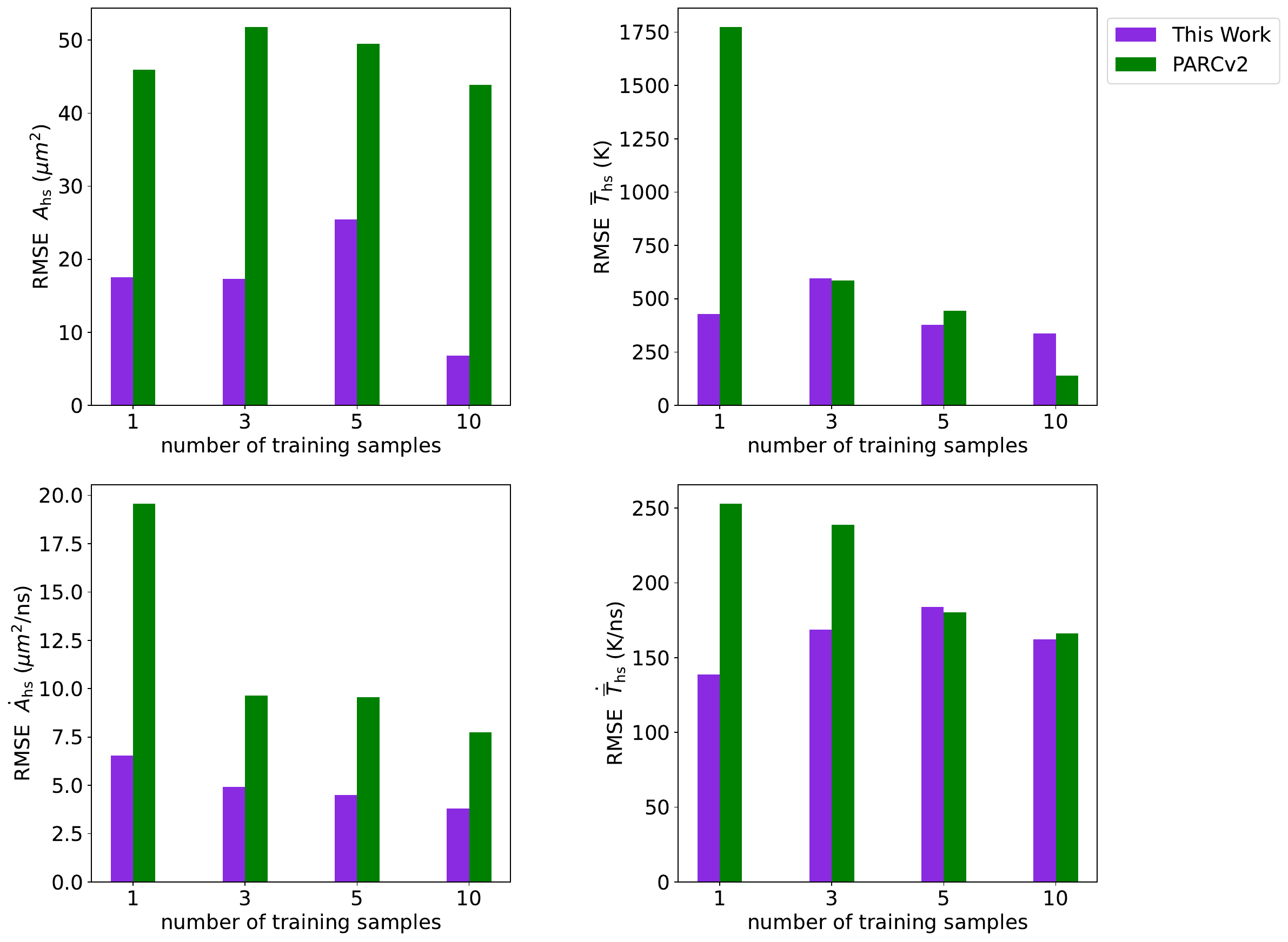}
    \caption{RMSE of the predicted sensitivity QoI by our model and PARCv2 in four training scenarios.}
    \label{fig:rmse_hotspot_metrics}
\end{figure}

Figure \ref{fig:LP_Pv2_hotspot_metrics} (a) and (b) illustrate the time evolution of the total hotspot area $A_\hs$ and the average hotspot temperature $\bar{T}_\hs$ together with their rate of change for our model and PARCv2, respectively, in comparison with the ground-truth (DNS). All of these quantities of interest are calculated based on the predicted evolution of the temperature field for test samples by the two models trained on 10, 5, and 3 meso-scale samples, respectively. In terms of total hotspot area $A_\hs$ and its rate of change $\dot{A}_\hs$, our model's prediction shows closer agreement with the ground-truth while PARCv2 tends to underpredict these two QoI. In addition, the overall trend of our model's predicted average hotspot temperature $\bar{T}_\hs$ is consistent with the ground-truth, although it is underpredicted by $ \sim  350 \, \mathrm{K}$. On the other hand, PARCv2's prediction of $\bar{T}_\hs$ can be divided into two parts based on the first and the second half of the total time-interval. Roughly speaking, PARCv2 underpredicts $\bar{T}_\hs$ for time-period $t \in [0, 7.5 \, \mathrm{ns}]$ while overpredicts $\bar{T}_\hs$ for $t \in [7.5 \, \mathrm{ns}, 15 \, \mathrm{ns}]$. The sharp increase in PARCv2 predicted $\bar{T}_\hs$ around $t = 7.5 \, \mathrm{ns}$ leads to considerable deviation of its predicted ${\dot{\bar{T}}}_\hs$ from ground-truth in that region. Figure \ref{fig:rmse_hotspot_metrics} shows the root mean squared error (RMSE) of the predicted quantities of interest by the two models trained in the aforementioned four scenarios. The RMSE of our model's predicted total hotspot area $A_\hs$ and its rate of change $\dot{A}_\hs$ is less than $\%50$ of the corresponding RMSE values for PARCv2 in all the training scenarios. On the other hand, the RMSE of PARCv2 predicted average hotspot temperature $\bar{T}_\hs$ decreases much faster than our model by increasing the number of training samples, leading to lower RMSE at the 10-sample training scenario.

\subsection{Uncertainty quantification for model prediction}

There are two main sources of uncertainty in the predictions made by our model: \textit{i)} the stochasticity due to various morphologies of the input microstructures; \textit{ii)} the stochastic nature of the learned dynamics by our model over the latent space. A comprehensive investigation of these two sources of stochasticity and their relation to each other is an important subject that is beyond the scope of this paper, and we leave it for future work. In this subsection, we focus on the second source of stochasticity as mentioned above, \textit{i.e.}, the uncertainty due to the random sampling made by the variational decoders in our model for predicting the physical fields of interest. We consider the following setup in order to evaluate the uncertainty in our model's predicted temperature field and sensitivity QoI for the problem of shock-induced energy localization in energetic material. Given the fixed initial condition for all the test samples, we run the model's inference for 50 trials based on the full \emph{stochastic} dynamics over the latent space, \textit{i.e.}, the evolution of the latent mean field $\bar{Z}$ and the latent log-variance field $Z^\sigma$. We compute the pixel-wise mean and standard deviation of the resulting ensemble of the predicted temperature field $T$ evolution. In addition, for each test sample, we run the model's inference based on the \emph{deterministic} dynamics over the corresponding slow manifold, \textit{i.e.}, \emph{only} the evolution of the latent mean field $\bar{Z}$.

\begin{figure}
     \centering
     \begin{subfigure}{\textwidth}
         \centering
         \includegraphics[width=\textwidth]{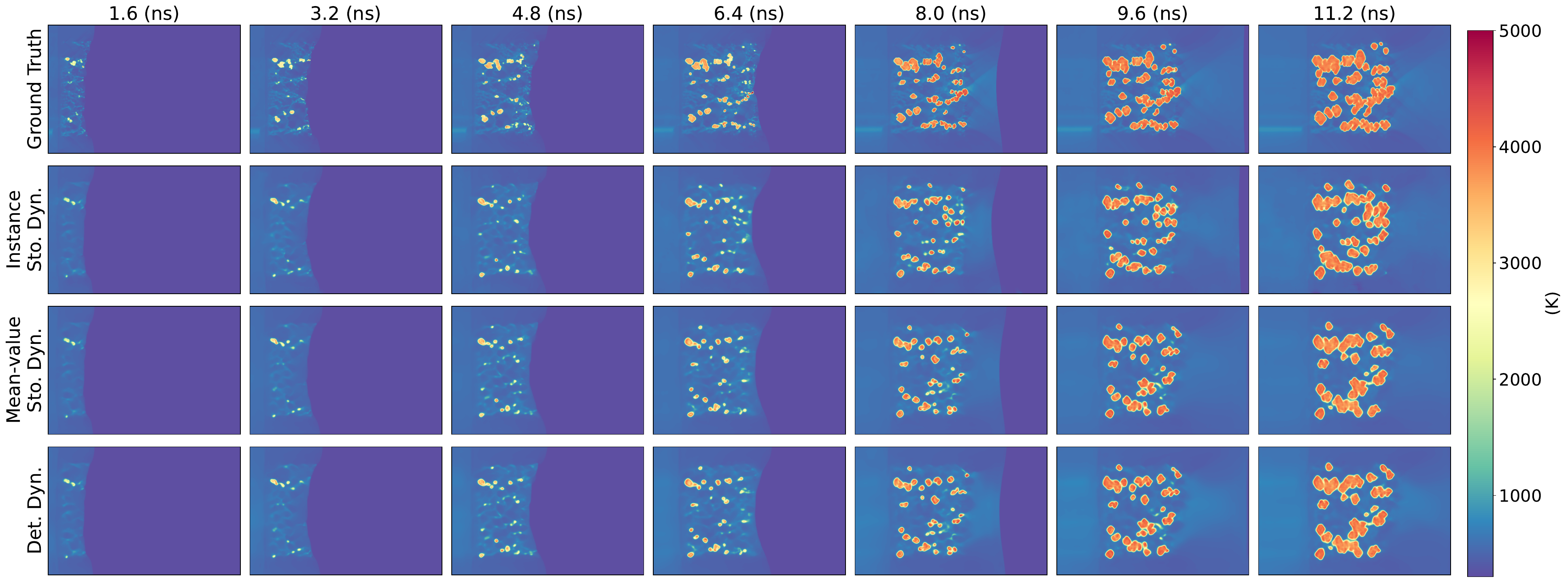}   
     \end{subfigure}
     \begin{subfigure}{\textwidth}
         \centering
         \includegraphics[width=\textwidth]{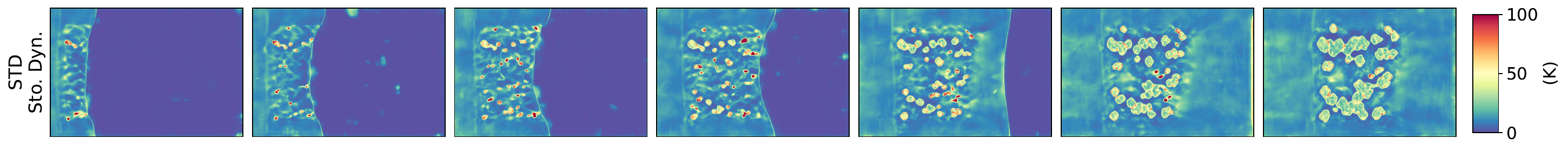}
     \end{subfigure}
     \caption{Predicted temperature field evolution based on the stochastic and deterministic dynamics over the latent space.} 
     \label{fig:UQ_T_pred_mean_var}
\end{figure}

\begin{figure}
    \centering
    \includegraphics[width=0.9\linewidth]{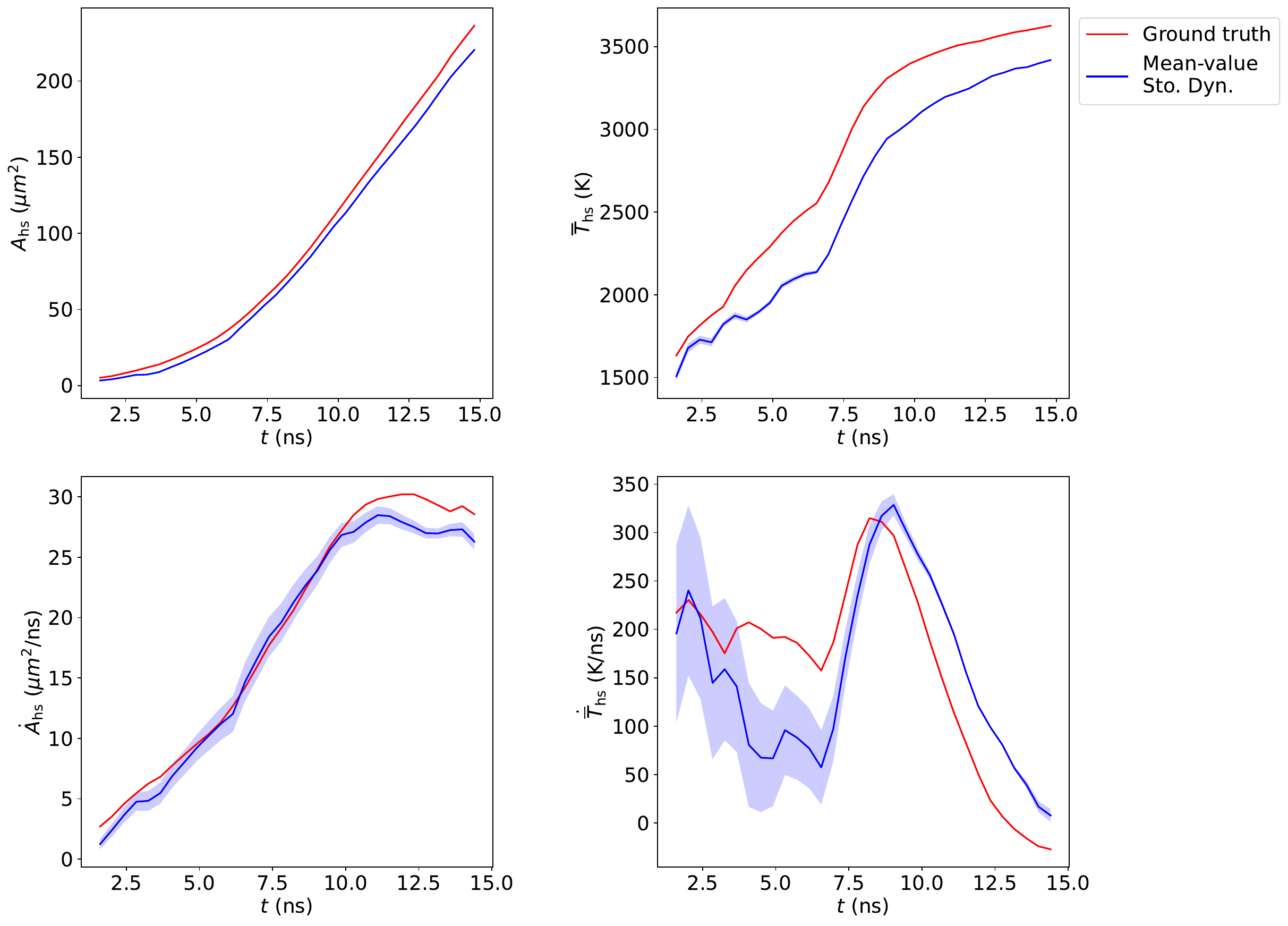}
    \caption{Mean-value and standard deviation of the predicted sensitivity QoI.}
    \label{fig:UQ_hotspot_metrics}
\end{figure}

Figure \ref{fig:UQ_T_pred_mean_var} illustrates the predicted temperature field $T$ evolution corresponding to the stochastic and deterministic inference of our model in comparison with the ground-truth for a test sample. The second row shows an instance of stochastic dynamics-based prediction, \textit{i.e.}, a sample from the 50-trial ensemble. The third and fifth rows show the pixel-wise mean-value and standard deviation of the corresponding 50-trial ensemble, respectively. The fourth row illustrates the predicted temperature evolution based on the deterministic dynamics over the latent space. As expected, the mean-value of stochastic dynamics-based predictions is quite close to the deterministic dynamics-based prediction. The standard deviation of predicted $T$ in the growth stage of hotspots is $\sim 50 \, \mathrm{K}$. However, in the ignition stage, we see standard deviation up to $\sim 150 \, \mathrm{K}$, which happens in false positively-predicted ignitions. Figure \ref{fig:UQ_hotspot_metrics} shows the mean-value and standard deviation of the predicted sensitivity QoI associated with the stochastic dynamics-based 50-trial ensemble, which has been averaged over all the test samples.   
 
\section{Discussion} \label{sec:discussion}

This work presented a meta-learning framework for learning the meso-scale thermomechanical behavior of shock-loaded porous energetic material (EM) in a scarce meso-scale data regime by leveraging inexpensive micro-scale simulations. In this framework, it was proposed to consider the corresponding meso-scale dynamics as a composition of interacting building blocks whose dynamics can be learned from micro-scale physics. A probabilistic autoregressive model was suggested to learn a reduced representation of the micro-scale dynamics by capturing the slow manifold associated with the evolution of the temperature and pressure fields in shock-induced single pore collapse in EM. The proposed structure for the autoregressive functional of our model encapsulates the short-term decoupled evolution of the involved physical fields together with their correlated long-term evolution. It was shown that, after learning a latent representation of the micro-scale dynamics, our model can be trained, relatively quickly, on a small dataset of meso-scale simulations in order to learn the higher-order spatiotemporal interactions between the physical fields at meso-scale. Because of the fully convolutional, \textit{i.e.}, CNN-based, architecture of the autoregressive functional, our model is able to learn the correlation between the neighboring micro-scale building blocks of the meso-scale physics. 

The predictive capability of our model was compared with a physics-aware recurrent convolutional neural network (PARC) trained only on the same small meso-scale dataset. The results showed that our model could outperform PARC in terms of predicting the formation and advection of localized high temperature regions (hotspots), in addition to the predicted temperature values for the hotspots. Also, the prediction of our model and PARC was compared against the ground-truth numerical simulations in terms of several EM sensitivity metrics which include total hotspot area, average hotspot temperature, and their rate of change over time. The comparison between the prediction performance of our model and PARC trained on the same meso-scale dataset showed that micro-scale dynamics contains crucial information that can considerably help a deep learning-based model to learn the meso-scale physics underlying EM thermo-mechanics.  

For future work, we plan to investigate the possibility of performance improvement of our proposed model by considering other neural network architectures, other than the U-Net architecture considered in this work, for the decoupled and the correlation autoregressive functionals. In particular, we expect that the performance of the model can be improved by considering an architecture with hidden states, \textit{e.g.}, Convolutional LSTM \citep{shi2015convolutional}, which can enhance the ability of the model in learning the long-term temporal correlations. Additionally, the trade-off between the amount of micro- and meso-scale data which is necessary for training our model needs to be investigated and optimized towards the least possible computational cost in overall. Finally, although the scope of the present work was limited to the thermomechanical behavior of energetic materials, the applicability of the proposed framework to other multiphysics problems with multiscale spatiotemporal dynamics can be a direction of future research.

\section*{Acknowledgments}
This work was supported by the National Science Foundation under Grant No. DMREF-2203580.

\section*{Author Declarations}
The authors declare that they have no known competing financial interests or personal relationships that could have appeared to influence the work reported in this paper.

\appendix
\section{Appendix}

\subsection{Details on Implementation and Training} \label{sec:appendix_implementation}

The encoder of the variational autoencoder in our model has a VGG-Net \citep{simonyan2014very} architecture with no fully-connected layers. It consist of four convolutional blocks with filter size 3, stride 1, rectified linear unit (ReLU) activation function, and number of channels equal to [64, 128, 256, 512], respectively. The output of each convolutional block is down-sampled by a $2 \times 2$ max-pooling operation with stride 2 to reduce the spatial dimensions by half and, hence, double the size of receptive field. The final feature map with 512 channels is processed through two subsequent convolutional layers with filter size 3, stride 1, no activation function, and number of channels equal to 4, which result in the associated latent-mean and latent-log-variance fields. The decoder has a symmetric architecture to the encoder with the exception that the max-pooling operation is replaced with $2 \times 2$ up-sampling operation.

All the autoregressive functionals in our model have a U-Net \citep{ronneberger2015u} architecture. It consist of a contracting path and an expansive path. The contractive path consist of three convolutional blocks with filter size 3, stride 1, ReLU activation function, and number of channels equal to [64, 128, 256], respectively, where each convolutional block is followed by a $2 \times 2$ max-pooling operation of stride 2 for down-sampling. The expansive path consist of three convolutional blocks symmetric to the contractive path where the max-pooling operation is replaced with $2 \times 2$ up-sampling operation followed by a concatenation with corresponding high-resolution feature map from the contractive path.

Our model is trained using the ADAM optimizer \citep{kingma2015adam}. The variational autoencoder corresponding to each physical field is trained for 500 epochs with learning rate $10^{-5}$. The autoregressive latent evolution module is trained following a \emph{curriculum learning} \citep{soviany2022curriculum} strategy in which the predicted time horizon increases successively from 1 to 4. Our model is trained for $[400, \, 300, \, 300, \, 200]$ epochs with learning rates $[10^{-4}, \, 8 \times 10^{-5}, \, 6 \times 10^{-5}, \, 4 \times 10^{-5}]$ according to the above-mentioned four-step curriculum learning.

In our experiments, we maintained the same architecture of PARCv2 as presented in \citep{nguyen2025parcv2}. We trained the PARCv2 differentiator with ADAM optimizer following the same curriculum learning setup which was considered for training of the latent evolution module in our model. We did not consider a data-driven integrator for PARCv2 in this experiment and only used the numerical integrator RK4.   

\subsection{Numerical Simulation Data} \label{sec:appendix_DNS}

The micro- and meso-scale simulations of the shock-induced energy localization in energetic material considered in this paper are generated using an in-house multi-material flow solver SCIMITAR3D \citep{rai2015mesoscale,rai2017high,rai2018three}. Details of its numerical framework for computing the reactive shock dynamics in energetic materials are discussed extensively in several publications \citep{kapahi2013dynamics,rai2017collapse,das2020sharp}. Also, its has been validated against experiment \citep{rai2020evaluation} and molecular dynamics simulations \citep{das2021molecular} for high-speed multi-material shock and impact problems.

We use a dataset containing 119 instances of single-pore collapse in addition to 13 instances of meso-scale simulations of shock-initiated reaction in pressed energetic material (Class V HMX \citep{molek2017microstructural}). For each simulation instance, the microstructural sample is loaded with a shock applied at the left boundary of the domain with the pressure loading of 9.5 GPa. The shock then traverses through the microstructure from the left to right. The single-pore collapse (\textit{i.e.}, micro-scale) instances have the spatial dimension of $1.5 \times 2.25 ({\mu m}^2)$ with void diameters ranging from $0.15 (\mu m)$ to $1.5 (\mu m)$. To capture the intricate details of void collapse and hotspot shapes, the micro-scale simulations employed a range of grid sizes from $0.7 (nm)$ to $5 (nm)$. This resulted in computational meshes with approximately $600 \times 1200$ grid points. The choice of grid size was informed by previous mesh convergence studies on void collapse \citep{rai2017high} which indicated that each void should be covered by a minimum of around 200 grid cells to ensure accurate representation (see \citep{nguyen2022multi} for details). The calculated temperature $T$, pressure $P$ and microstructural morphology $\mu$ fields in the micro-scale simulations were recorded at equal time intervals resulting in 19 equally-spaced snapshots with $\Delta t = 0.17 (ns)$.\footnote{In the numerical simulation, in order to capture the sharp interface between voids and the HMX substrate, the interfaces are modeled using the narrow band levelset approach \citep{sethian1996Level,nguyen2022multi}. However, the microstructural morphology field $\mu$, considered in this paper for training the neural network models, is a binarized version of the levelset representation.} For training the micro-scale dynamics model, the $(T, P, \mu)$ snapshots corresponding to the single-pore collapse are resolved with a uniform grid of size $160 \times 256$ pixels. 

The meso-scale microstructure samples have the spatial dimension of $25 \times 25 ({\mu m}^2)$ padded into a homogeneous HMX material domain of size $35 \times 51 ({\mu m}^2)$. In order to capture the same level of physical detail at meso-scale simulations, one needs to use similar grid sizes to the single-pore collapse simulations. However, this leads to computational meshes that are hundreds of times larger than those used in micro-scale simulations. Consequently, this places a significant burden on computational resources, including increased memory usage, longer processing times, and higher computational power requirements (see \citep{seshadri2022meso} for details). The calculated evolution of $(T, P, \mu)$ fields in the meso-scale simulations were recorded at equal time intervals resulting in 38 equally-spaced snapshots with $\Delta t = 0.4 (ns)$ and resolved with a uniform grid of size $320 \times 480$ pixels. 

\subsection{Comparison of Prediction Results for More Test Samples} \label{sec:appendix_pred_test}

\begin{figure}
     \centering
     \begin{subfigure}{\textwidth}
         \centering
         \includegraphics[width=\linewidth]{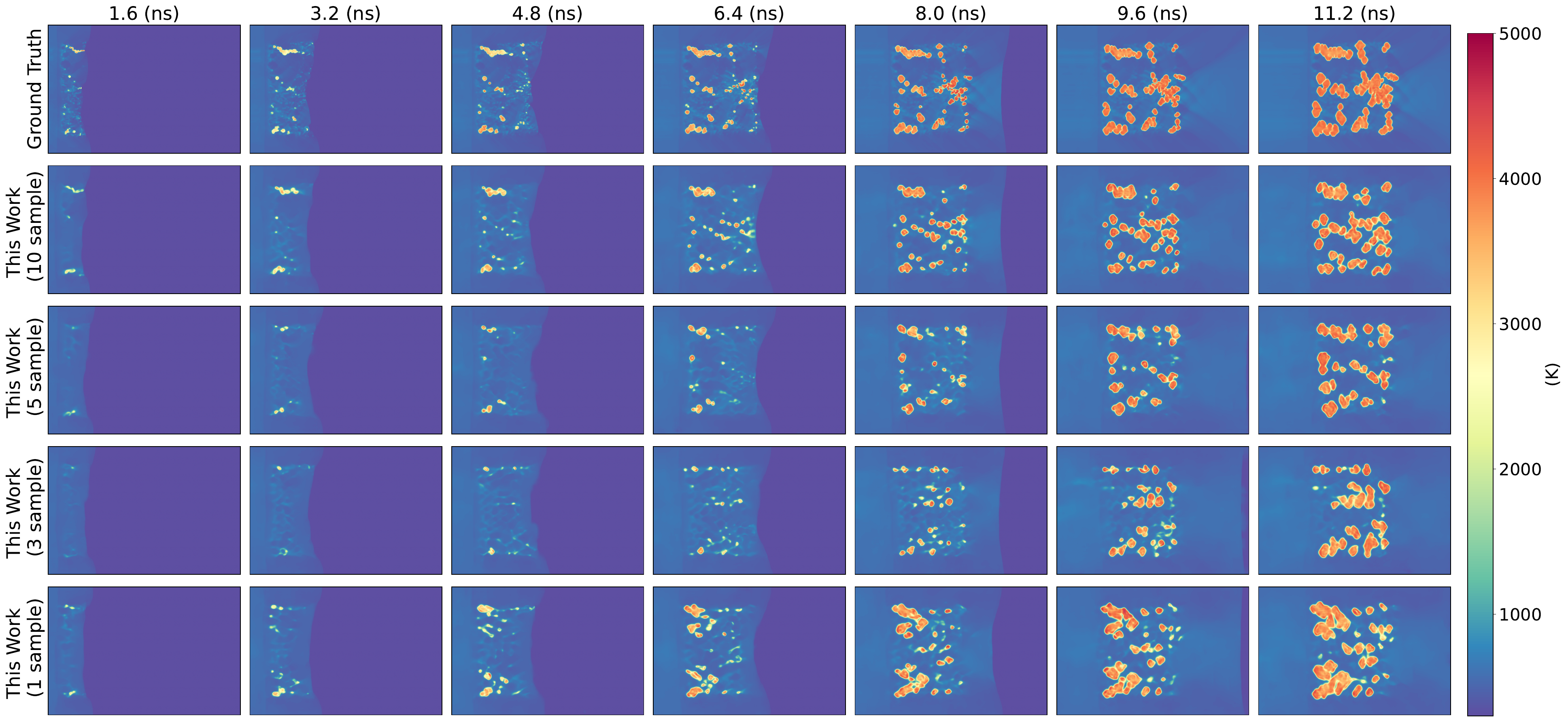}
         \caption{}         
     \end{subfigure}
     \begin{subfigure}{\textwidth}
         \centering
         \includegraphics[width=\linewidth]{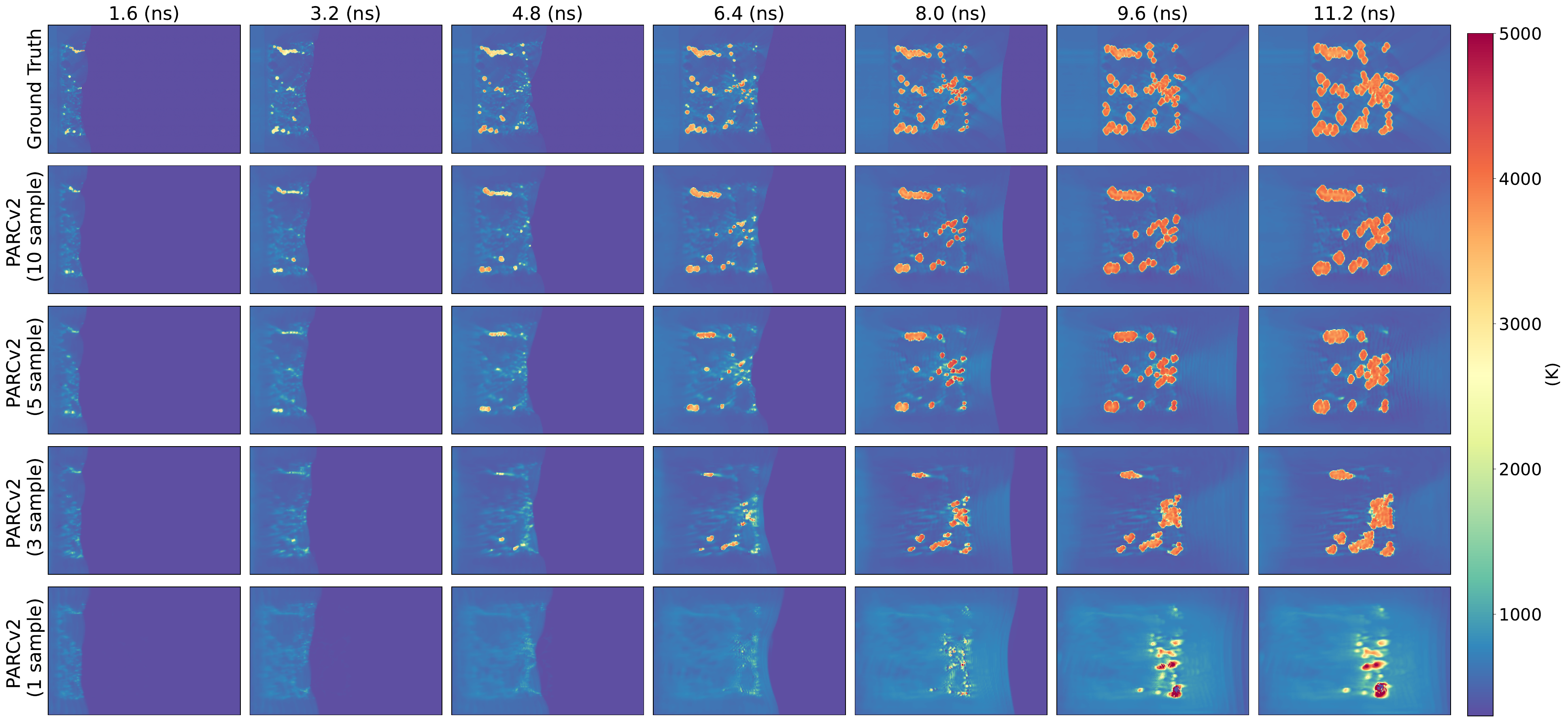}
         \caption{}
     \end{subfigure}
     \caption{Test sample \#2: Comparison between the predicted temperature field evolution by (a) our model and (b) PARCv2 in four training scenarios.}    
     \label{fig:LP_Pv2_comparison_T_prediction_2}
\end{figure}

\begin{figure}
     \centering
     \begin{subfigure}{\textwidth}
         \centering
         \includegraphics[width=\linewidth]{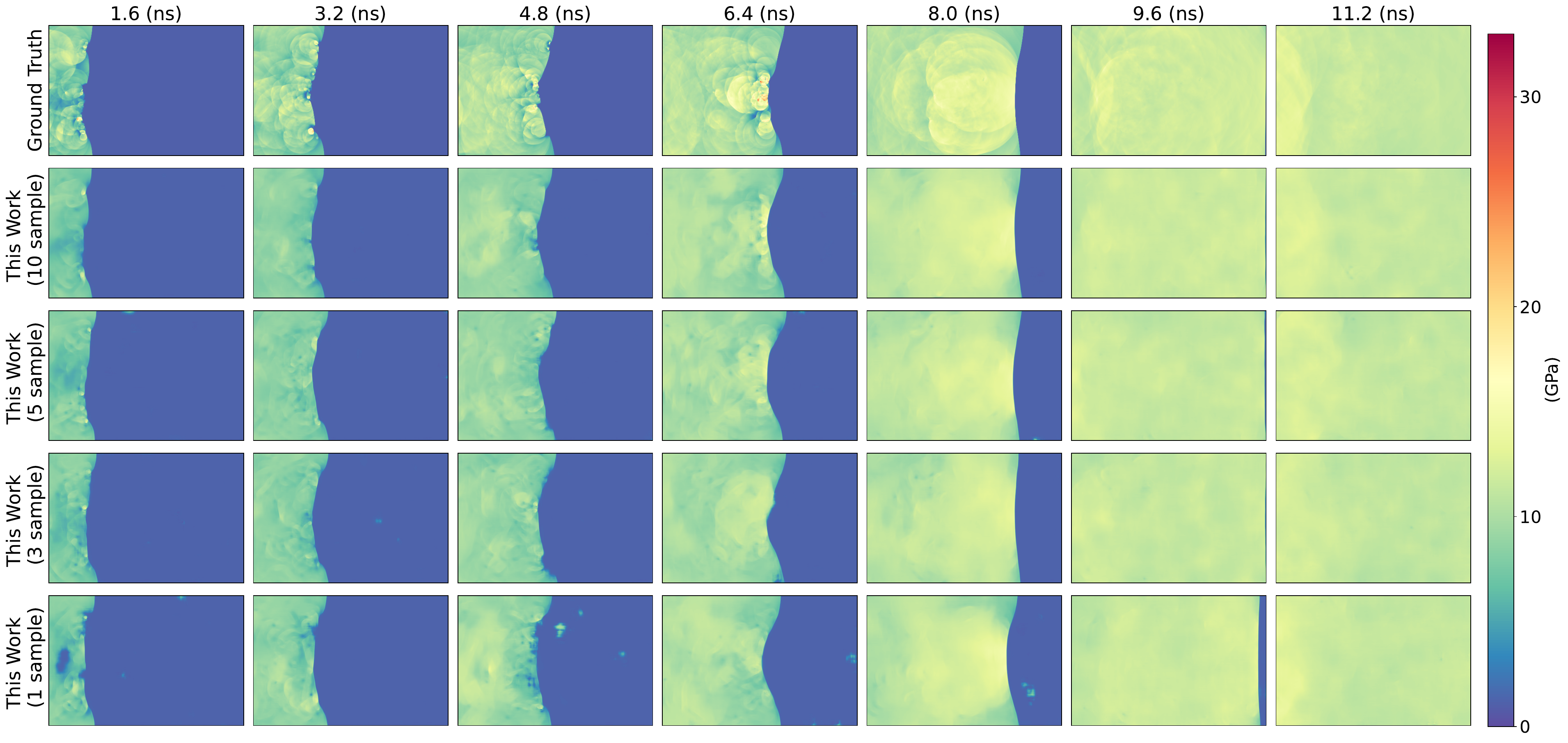}
         \caption{}         
     \end{subfigure}
     \begin{subfigure}{\textwidth}
         \centering
         \includegraphics[width=\linewidth]{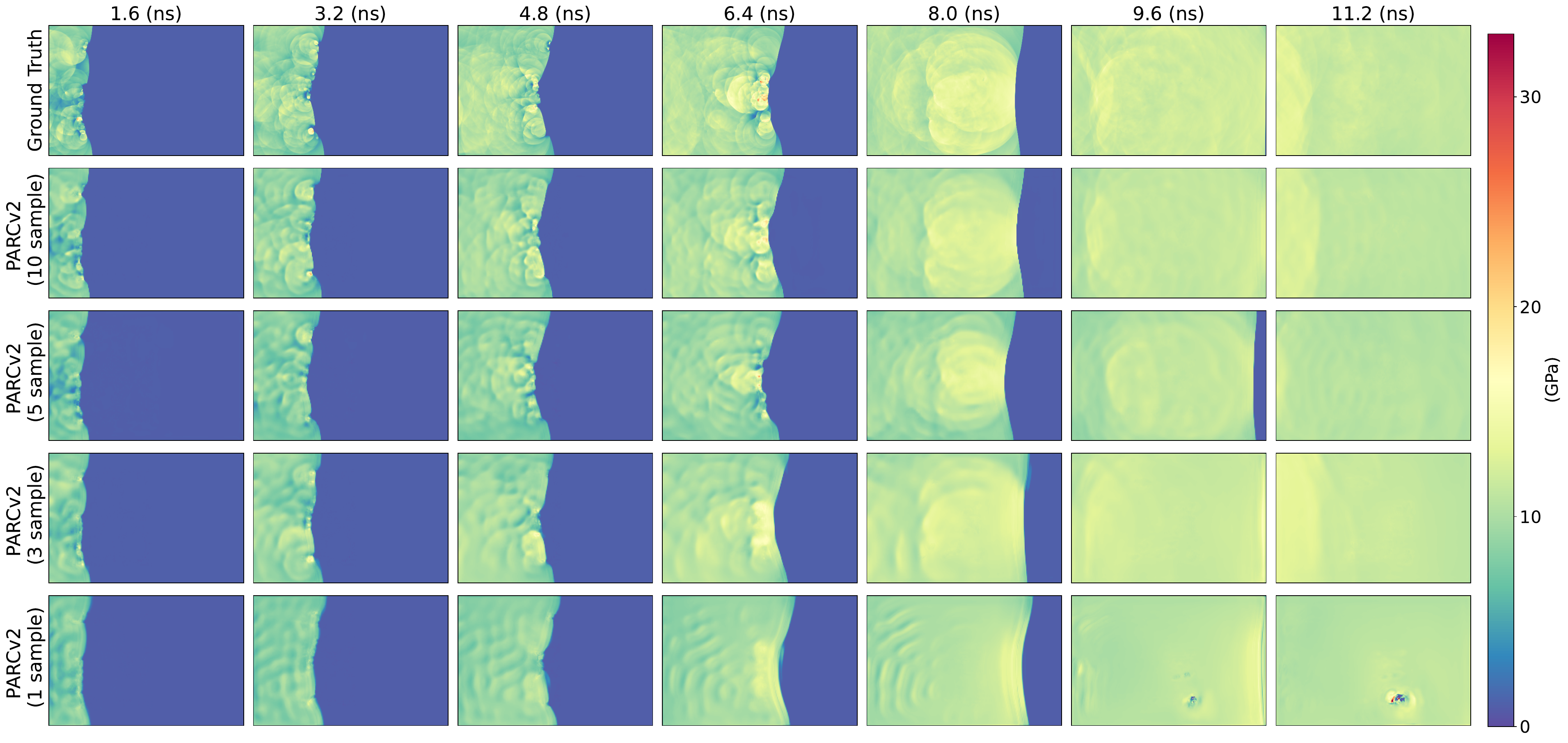}
         \caption{}
     \end{subfigure}
     \caption{Test sample \#2: Comparison between the predicted pressure field evolution by (a) our model and (b) PARCv2 in four training scenarios.}    
     \label{fig:LP_Pv2_comparison_P_prediction_2}
\end{figure}

\begin{figure}
     \centering
     \begin{subfigure}{\textwidth}
         \centering
         \includegraphics[width=\linewidth]{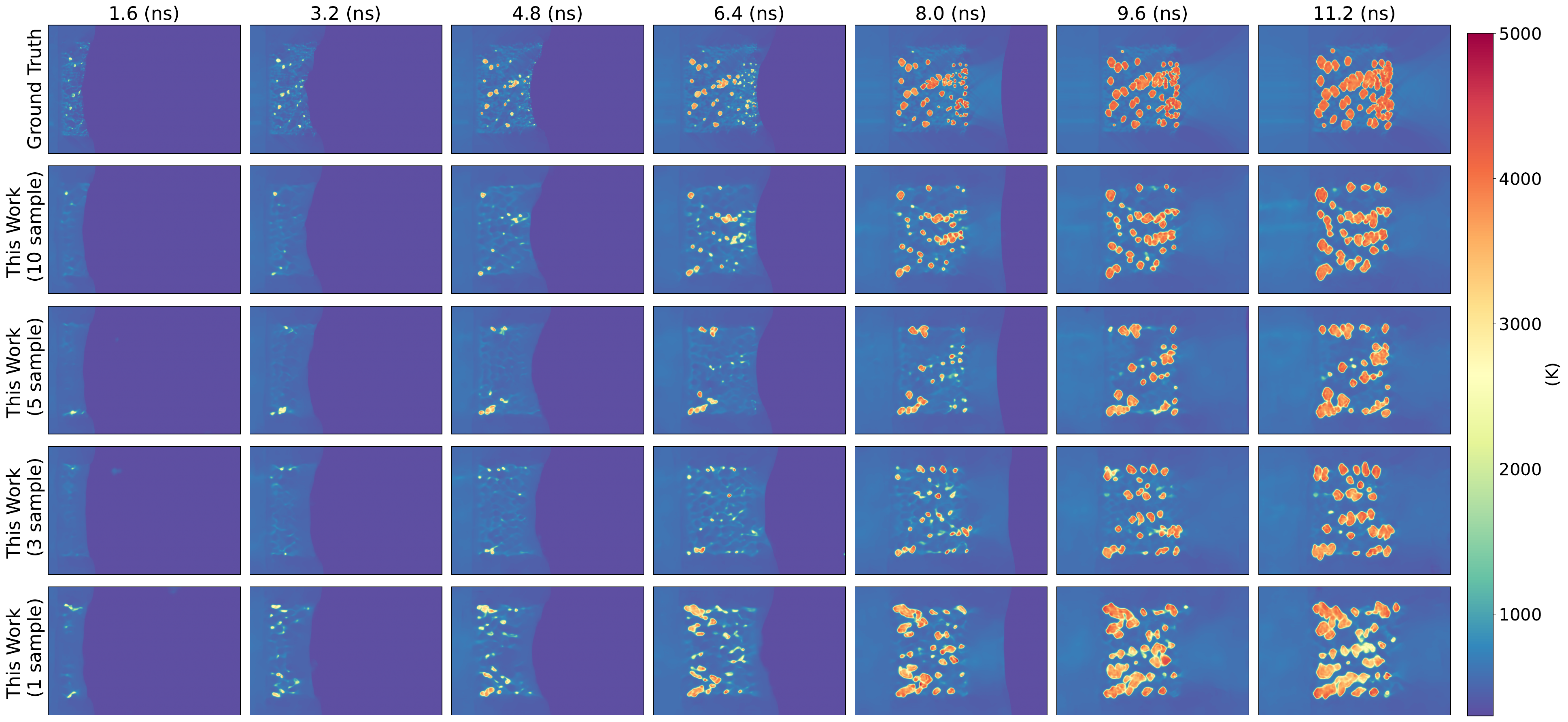}
         \caption{}         
     \end{subfigure}
     \begin{subfigure}{\textwidth}
         \centering
         \includegraphics[width=\linewidth]{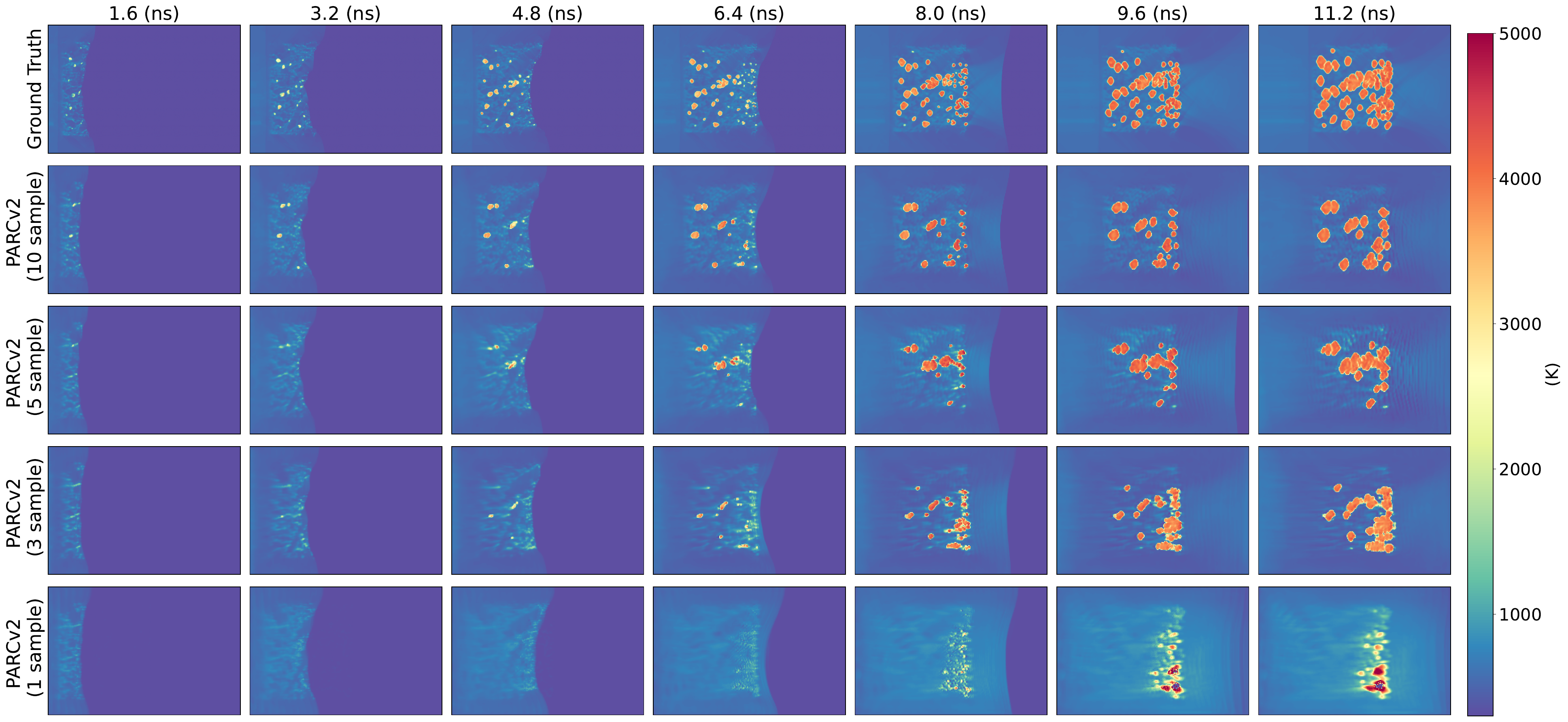}
         \caption{}
     \end{subfigure}
     \caption{Test sample \#3: Comparison between the predicted temperature field evolution by (a) our model and (b) PARCv2 in four training scenarios.}    
     \label{fig:LP_Pv2_comparison_T_prediction_3}
\end{figure}

\begin{figure}
     \centering
     \begin{subfigure}{\textwidth}
         \centering
         \includegraphics[width=\linewidth]{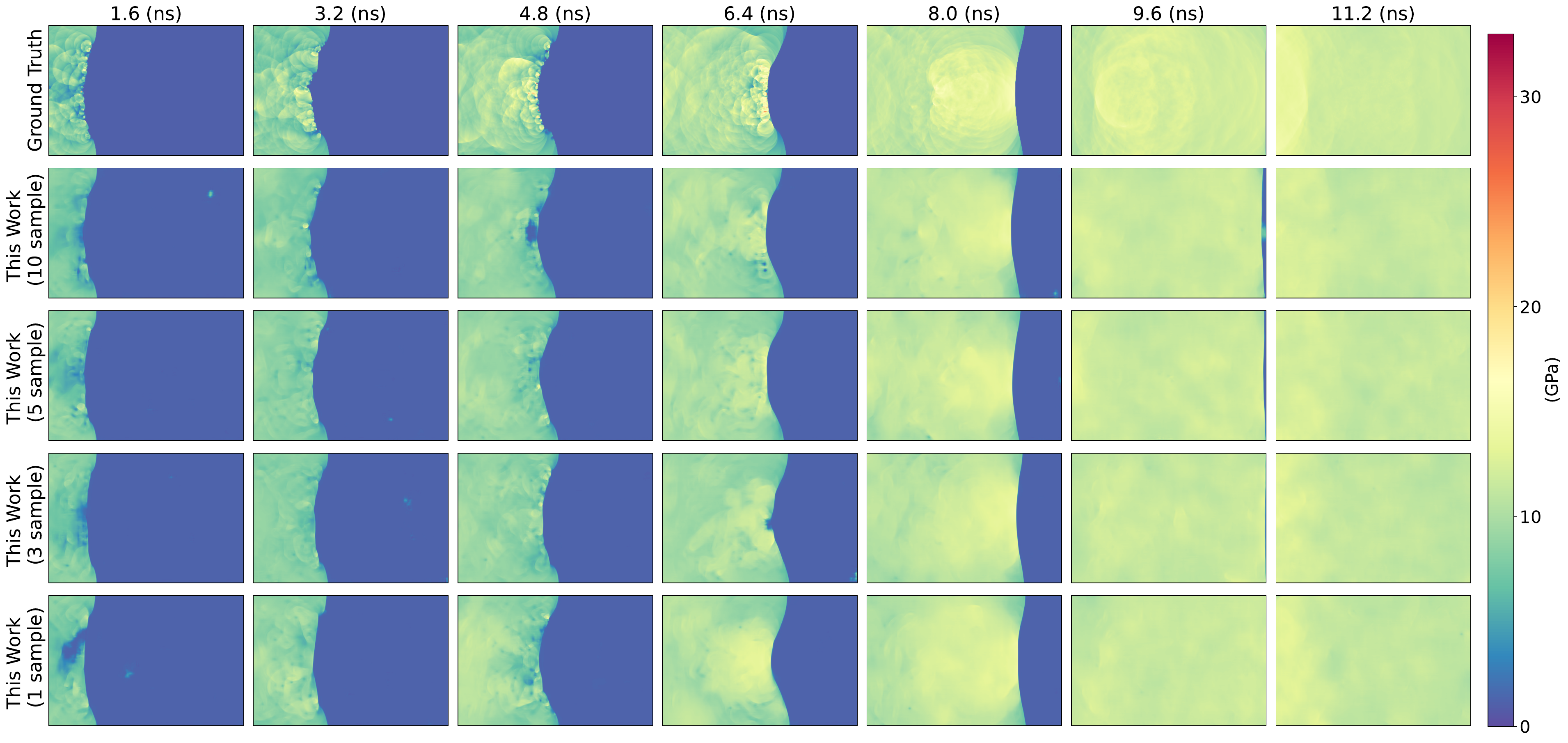}
         \caption{}         
     \end{subfigure}
     \begin{subfigure}{\textwidth}
         \centering
         \includegraphics[width=\linewidth]{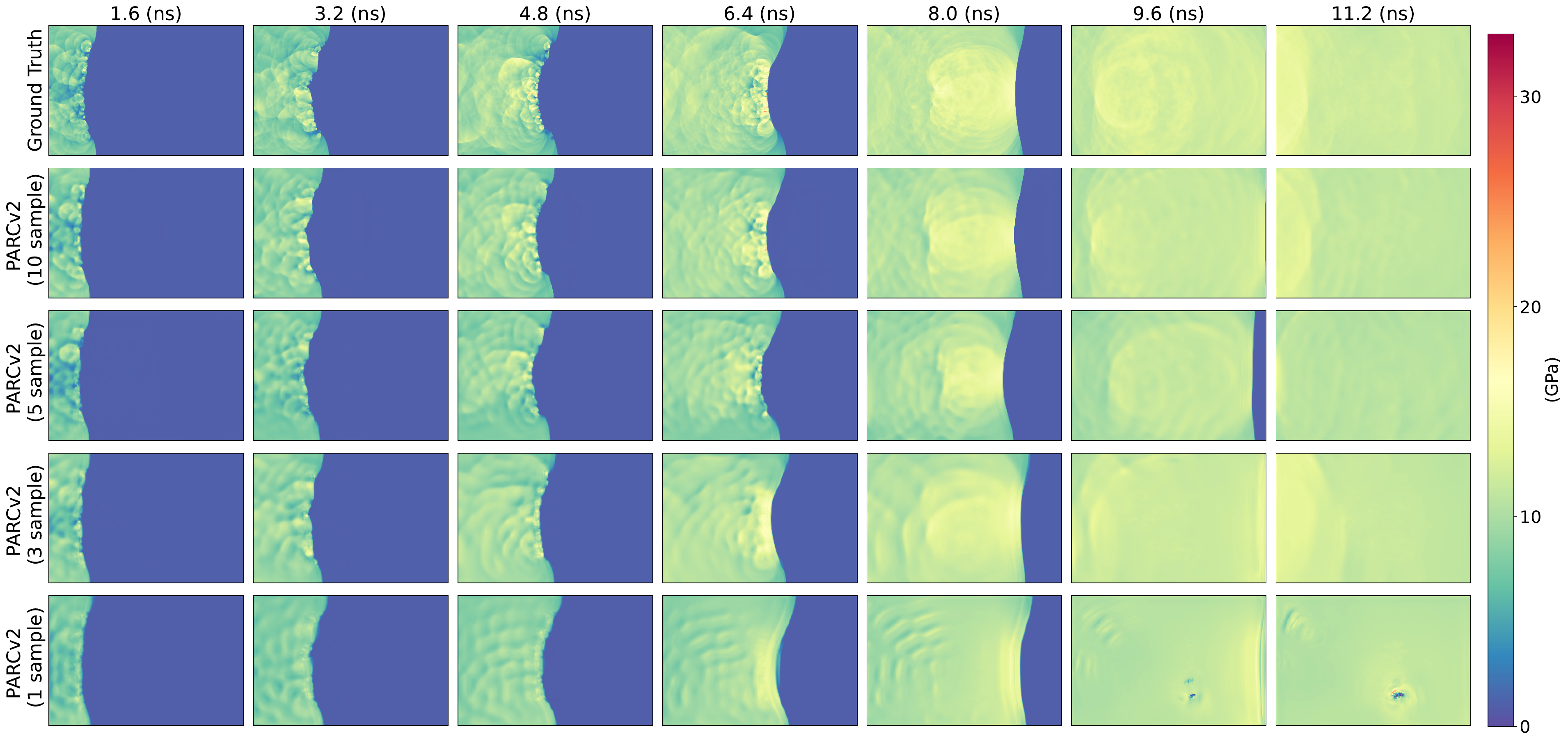}
         \caption{}
     \end{subfigure}
     \caption{Test sample \#3: Comparison between the predicted pressure field evolution by (a) our model and (b) PARCv2 in four training scenarios.}    
     \label{fig:LP_Pv2_comparison_P_prediction_3}
\end{figure}

In order to make a more comprehensive comparison between the predicted evolution of the temperature $T$ and pressure $P$ fields by our model and PARCv2 in the four training scenarios discussed in Section \ref{sec:compare_parc_qualitative}, we present the prediction results for all the other test samples in Figures \ref{fig:LP_Pv2_comparison_T_prediction_2} - \ref{fig:LP_Pv2_comparison_P_prediction_3}.

\bibliographystyle{unsrt} 

\end{document}